\documentclass[11pt]{article}

\usepackage{fontspec}
\usepackage{amsmath}
\usepackage{unicode-math}
\usepackage{microtype}
\usepackage[margin=1in]{geometry}
\usepackage{graphicx}
\usepackage{booktabs}
\usepackage{tabularx}
\usepackage{multirow}
\usepackage{siunitx}
\usepackage{xcolor}
\usepackage[authoryear,round]{natbib}
\usepackage[
  colorlinks=true,
  linkcolor=blue!60!black,
  citecolor=blue!60!black,
  urlcolor=blue!60!black
]{hyperref}
\newcommand{\doi}[1]{doi: \href{https://doi.org/#1}{\nolinkurl{#1}}}
\usepackage[capitalise,noabbrev]{cleveref}
\usepackage{caption}
\usepackage{subcaption}
\usepackage{enumitem}
\usepackage{float}
\usepackage{needspace}

\graphicspath{{figures/}}
\setlist{nosep}

\newcommand{\datafigure}[5][!htbp]{%
  \begin{figure}[#1]
    \centering
    \includegraphics[width=#3]{#2.pdf}
    \caption{#4}
    \label{#5}
  \end{figure}}

\newcommand{\NumTenProducts}{12}

\newcommand{\NumTenBrewPostman}{139{,}258}
\newcommand{\NumTenBrewBruno}{75{,}820}
\newcommand{\NumTenBrewBrunoPctPostman}{54}
\newcommand{\NumTenBrewNotion}{60{,}782}

\newcommand{\NumTenBrewAppFlowyPctNotion}{2.4}

\newcommand{\NumTenBrewPostmanFirst}{63{,}350}
\newcommand{\NumTenBrewNotionFirst}{6{,}140}

\newcommand{\NumTenContribPre}{167}
\newcommand{\NumTenContribPost}{4}

\newcommand{\NumCapMetrFitDoublingDays}{129}

\newcommand{\NumCapMetrFitBootCiLow}{102}
\newcommand{\NumCapMetrFitBootCiHigh}{154}
\newcommand{\NumCapMetrFitCiPercent}{95}

\newcommand{\NumCapMetrFitGrowthPerYear}{7.1}

\newcommand{\NumCapMetrTopPFifty}{17.4 hours}

\newcommand{\NumCapMetrTopPEighty}{3.1 hours}
\newcommand{\NumCapMetrReliableMaxHours}{16}
\newcommand{\NumCapMetrTopDate}{April 2026}
\newcommand{\NumCapMetrGptFourPFifty}{4 minutes}
\newcommand{\NumCapMetrLastRelease}{April 2026}

\newcommand{\NumCapEciLatestMonth}{September 2026}
\newcommand{\NumCapEciRecentN}{12}

\newcommand{\NumCapFrontierSweTrials}{5}

\newcommand{\NumCapFrontierSweTopScore}{65.5}

\newcommand{\NumCapAiderCostSixtyFirst}{186.50}
\newcommand{\NumCapAiderCostSixtyFirstMonth}{December 2024}
\newcommand{\NumCapAiderCostSixtyLast}{4.80}
\newcommand{\NumCapAiderCostSixtyLastMonth}{August 2025}
\newcommand{\NumCapAiderCostSixtyFold}{39}

\newcommand{\NumAicTotalLastMonthM}{16.46}

\newcommand{\NumAicTotalFoldYoY}{19.8}

\newcommand{\NumAicClaudeShareOfAILastMonth}{87}

\newcommand{\NumAicStockClaudeMd}{872{,}448}
\newcommand{\NumAicStockAgentsMd}{1{,}105{,}920}

\newcommand{\NumAltMinStars}{20}
\newcommand{\NumAltNarrowMinStars}{50}
\newcommand{\NumAltCandidates}{1{,}211}
\newcommand{\NumAltIncluded}{575}
\newcommand{\NumAltIncludedAgentEra}{191}
\newcommand{\NumAltAgentEraPct}{33}

\newcommand{\NumAltNarrowTotal}{45}
\newcommand{\NumAltAiFilesPctAgentEra}{47}
\newcommand{\NumAltAiFilesPctBefore}{11}
\newcommand{\NumAltScreenStudioBroad}{6}

\newcommand{\NumAltSmallpdfBroad}{5}
\newcommand{\NumAltSmallpdfAgentEra}{4}

\newcommand{\NumBeaSharePrepackagedLast}{49.6}

\newcommand{\NumBeaSharePrepackagedTwentyTwentyTwo}{46.3}

\newcommand{\NumBeaShareOwnAccountLast}{13.3}

\newcommand{\NumBeaShareOwnAccountTwentyTwentyTwo}{14.0}

\newcommand{\NumCraftApps}{12}
\newcommand{\NumCraftCreatedFirst}{2026-09-30}
\newcommand{\NumCraftCreatedLast}{2026-10-07}

\newcommand{\NumCraftHistoryDays}{12.0}

\newcommand{\NumCraftCommitsTotal}{11{,}760}

\newcommand{\NumCraftAiAnyPct}{61}

\newcommand{\NumCraftLocTotal}{2{,}633{,}601}

\newcommand{\NumCraftTestFns}{30{,}928}

\newcommand{\NumCraftDownloadsTotal}{4{,}145{,}191}

\newcommand{\NumCraftIssueReportersNonMember}{2{,}075}

\newcommand{\NumCraftPrAuthorsNonMember}{876}

\newcommand{\NumCraftReadyMedian}{50}

\newcommand{\NumBuildSurveyForegonePct}{32}
\newcommand{\NumBuildSurveyN}{1,719}
\newcommand{\NumBuildSurveyNations}{97}
\newcommand{\NumBuildSurveyFieldStart}{May 4, 2026}
\newcommand{\NumBuildSurveyFieldEnd}{June 8, 2026}

\newcommand{\NumVcbSelectedN}{9}

\newcommand{\NumVcbSubstepPassPct}{90}
\newcommand{\NumVcbLimitHours}{5}
\newcommand{\NumVcbLimitTurns}{1,000}
\newcommand{\NumVcbTopScore}{92.39}
\newcommand{\NumVcbLowSelectedScore}{86.17}

\newcommand{\NumFreeFiverrBuyersDeclinePct}{21.9}
\newcommand{\NumFreeFiverrMarketplaceRevenueDeclinePct}{15.5}

\newcommand{\NumFreeUpworkClientsDeclinePct}{4.1}
\newcommand{\NumFreeUpworkGsvDeclinePct}{3.6}
\newcommand{\NumFreeUpworkEbitdaGrowthPct}{12.3}

\newcommand{\NumGhtRepoAllGrowthTwelveM}{67}

\newcommand{\NumGhtOssAltRateTwentyTwo}{3.3}
\newcommand{\NumGhtOssAltRateLastTwelveM}{11.9}
\newcommand{\NumGhtOssAltSTenRateTwentyTwo}{91}
\newcommand{\NumGhtOssAltSTenRateLastTwelveM}{290}
\newcommand{\NumGhtSelfhostAltRateTwentyTwo}{0.4}
\newcommand{\NumGhtSelfhostAltRateLastTwelveM}{4.5}

\newcommand{\NumHnShowPreMean}{823}
\newcommand{\NumHnShowLastTwelveMean}{3,846}

\newcommand{\NumHnShowShareStoriesPre}{3.3}
\newcommand{\NumHnShowShareStoriesLastTwelve}{12.9}

\newcommand{\NumHnAltShareTwentyTwo}{1.50}
\newcommand{\NumHnAltShareLastTwelve}{1.25}

\newcommand{\NumIgPushesYoY}{80}
\newcommand{\NumIgPushesYoYPreAgentMean}{15}

\newcommand{\NumIgDevNetAddLast}{13.7}
\newcommand{\NumIgDevNetAddTwentyThreeTwentyFour}{5.9}

\newcommand{\NumIgPushEconomiesGrowingPct}{99}

\newcommand{\NumIgPushEconomiesN}{184}

\newcommand{\NumLabSoftDevLast}{78.3}

\newcommand{\NumLabSoftDevPeak}{233.9}
\newcommand{\NumLabSoftDevPeakDate}{February 2022}

\newcommand{\NumLabSoftDevAtChatGpt}{142.5}

\newcommand{\NumLabSoftDevSinceJanTwentyFivePct}{16.0}

\newcommand{\NumLabEmpSoftwarePublishersLast}{652.9}
\newcommand{\NumLabEmpSoftwarePublishersPeak}{664.3}
\newcommand{\NumLabEmpSoftwarePublishersPeakDate}{November 2022}

\newcommand{\NumLabCesLastMonth}{August 2026}

\newcommand{\NumPkgSensitivityFirstBreak}{August 2024}
\newcommand{\NumPkgSensitivityLastBreak}{August 2025}

\newcommand{\NumPkgChatgptBreak}{December 2022}
\newcommand{\NumPkgAgentBreak}{February 2025}
\newcommand{\NumPkgFilterWindowDays}{90}

\newcommand{\NumPkgPypiYoYPre}{26}

\newcommand{\NumPkgPypiYoYChatEra}{1}
\newcommand{\NumPkgPypiYoYAgentEra}{23}

\newcommand{\NumPkgCratesYoYPre}{39}

\newcommand{\NumPkgCratesYoYChatEra}{19}
\newcommand{\NumPkgCratesYoYAgentEra}{123}

\newcommand{\NumPkgPypiFilteredAgentSlope}{17.9}

\newcommand{\NumPkgCratesFilteredAgentSlope}{119.6}

\newcommand{\NumProjSuccessFifty}{50}
\newcommand{\NumProjSuccessEighty}{80}
\newcommand{\NumProjFastStartYear}{2023}
\newcommand{\NumProjSlowStartYear}{2019}
\newcommand{\NumProjMaxFitHours}{16}
\newcommand{\NumProjBootstrapSamples}{10{,}000}
\newcommand{\NumProjBandLowPercentile}{5}
\newcommand{\NumProjBandHighPercentile}{95}
\newcommand{\NumProjWorkdayHours}{8}
\newcommand{\NumProjWorkweekHours}{40}
\newcommand{\NumProjWorkmonthHours}{167}
\newcommand{\NumProjWorkyearHours}{2{,}000}

\newcommand{\NumProjFiftyFastWorkweekDate}{Dec 2026}

\newcommand{\NumProjFiftyFastWorkmonthDate}{Sep 2027}
\newcommand{\NumProjFiftyFastWorkmonthLow}{Apr 2027}
\newcommand{\NumProjFiftyFastWorkmonthHigh}{Feb 2028}

\newcommand{\NumProjFiftySlowWorkweekDate}{Nov 2027}

\newcommand{\NumProjFiftySlowWorkmonthDate}{Nov 2028}

\newcommand{\NumProjEightyFastWorkmonthDate}{Sep 2028}

\newcommand{\NumProjEightyFastWorkyearDate}{Jan 2030}

\newcommand{\NumProjEightySlowWorkmonthDate}{Aug 2030}

\newcommand{\NumProjEightySlowWorkyearDate}{Aug 2032}

\newcommand{\NumProjCheckpointEarlyYear}{2028}
\newcommand{\NumProjCheckpointMiddleYear}{2031}
\newcommand{\NumProjCheckpointLongYear}{2036}

\newcommand{\NumProjBacktestAboveN}{8}
\newcommand{\NumProjBacktestMedianRatio}{3.6}

\newcommand{\NumSoLastMonth}{2026-09}
\newcommand{\NumSoLastMonthQuestions}{986}

\newcommand{\NumSoBaselineTwentyTwoMonthly}{112,934}

\newcommand{\NumSoDeclineVsBaselinePct}{99.1}

\newcommand{\NumSoPreChatGPTDeclinePct}{36.7}

\newcommand{\NumSoSurveyReplacingVendors}{18.6}
\newcommand{\NumSoSurveyInternalTools}{63.0}
\newcommand{\NumSoSurveySmallScripts}{77.3}
\newcommand{\NumSoSurveyBuildN}{13,338}

\title{\textbf{The Customer Is the Builder:}\\[2pt]
\large How AI Coding Agents Are Rewriting the Economics of Software}

\author{Ripon Chandra Malo\thanks{Corresponding author: \href{mailto:ripon.malo@utah.edu}{ripon.malo@utah.edu}.}\\
\small University of Utah
\and Zimeng Yu\\
\small University of Utah
\and Tong Qiu\\
\small University of Utah}

\date{October 2026}

\begin{document}
\maketitle

\begin{abstract}
AI coding agents could change whether organizations buy software or build the
functionality they need. We focus on organizations and professional buyers whose
domain knowledge can help them specify and evaluate tools for their work. We define
the software substitution frontier as the functions customers could build and
maintain at an acceptable standard for less than buying them. Using public data,
we compare growing capacity to build with indicators of market change. Since
\NumProjFastStartYear{}, the length
of benchmark tasks that leading agents can complete at \NumProjSuccessFifty{}
percent success, measured in human working time, has doubled about every
\NumCapMetrFitDoublingDays{} days. A broad search
identified \NumAltIncluded{} potential open-source alternatives to selected
commercial products, although some functional matches still require independent
validation.
The Craft suite produced \NumCraftApps{} application repositories over
\NumCraftHistoryDays{} days of public development, but these remain early versions
with unverified completeness and reliability. Market evidence is mixed: purchased
software gained US investment share and tracked commercial installations grew.
Yet \NumBuildSurveyForegonePct{} percent of respondents to an organizational survey
reported forgoing at least one software purchase because their organization could
build the functionality with coding agents. These self-reports do not establish
the spending displaced or the reliability of the resulting tools. The gap motivates
a warning: customers' capacity to build could affect individual buying decisions
before a broad purchasing shift becomes visible. We examine this possibility using
benchmark projections and observable market tests. The projections are an
extrapolation, not a forecast of customer switching.
\end{abstract}

\section{Introduction}
\label{sec:intro}

Software companies have always competed with other software companies. They may
soon compete with their own customers. A firm that once needed a vendor could ask
a small internal team to build the
functionality it uses, with AI agents doing more of the programming. The first
version may be rough. Later versions may meet the firm's needs. Even before a
subscription is canceled, a credible option to build can change the value of buying.

By customers, we primarily mean organizations responsible for obtaining software:
companies, institutions and teams, including departments within large organizations.
Professional buyers, such as researchers and engineers purchasing tools for their
work, face the same choice. We call customers builders when they or their staff
create or meaningfully adapt a working tool for their own workflow. This does not require
reproducing an entire commercial product. It requires supplying the functions the
customer needs and accepting responsibility for their operation.

For many customers, that option was once too expensive. A company that needed a screen
recorder, a PDF editor or a project board bought a subscription because hiring developers
to build and maintain one would cost far more. Economists call this the make-or-buy
decision \citep{coase1937nature,williamson1975markets}. Coding agents may change its
answer, one workflow at a time. The customer may already understand the task, own
the relevant data and know how to check an answer. Agents could help turn that
knowledge into software without requiring a large development team. Existing
open-source components could further reduce the work needed to implement it.

Tools such as Claude Code and Codex can plan, write, test and revise software over many
steps. The length of benchmark tasks that frontier agents can finish has grown quickly
\citep{kwa2025measuring}. This makes the customer's ability to build a possible source of
competition for a vendor. Whether that ability becomes a dependable and cheaper option
depends on verification, maintenance, security and integration.

This idea is easy to state and easy to exaggerate. More code does not mean more shipped
products \citep{demirer2026writing}. More open-source alternatives do not mean fewer
paying customers. A careful randomized trial even found that early 2025 AI tools slowed
experienced developers down \citep{becker2025measuring}. So the question is empirical:

\begin{quote}
Is AI expanding the set of purchased software workflows that organizations and
professional buyers can economically build for themselves, and what would that mean
for developers and software companies?
\end{quote}

We split the argument into links that can fail separately. First, what can agents do?
Second, are people building more software? Third, does more of that software compete with
paid products? Fourth, are customers switching? The early links describe capacity and
creation. The last describes market decisions. \Cref{fig:chain} puts these
leading and lagging indicators side by side.

Our main finding is that the capacity to build software advances before a broad shift in purchasing becomes visible. Measures of agent capability and public software
creation rose sharply around early 2025, while the available market indicators show
mixed results. An organizational survey provides some evidence of changing
purchasing decisions: respondents reported forgoing software purchases in favor
of building \citep{customer_building_mckinsey2026}. However, the spending displaced
remains unmeasured. Together, these findings suggest that customer building could
affect individual purchasing decisions before its effects become visible in
aggregate market data. Stack Overflow illustrates how sharply use can fall when
alternatives become available: monthly questions declined by
\NumSoDeclineVsBaselinePct{} percent relative to the 2022 level, although the
decline began years before ChatGPT (\cref{fig:so}).

It illustrates how quickly use can change. It does not establish when or
whether paid software will follow.

This paper makes four contributions.
\begin{enumerate}
  \item We define the software substitution frontier, a framework for evaluating
        when customers can build and maintain the functionality they need for less
        than buying it (\cref{sec:framework}).
  \item We assemble a reproducible evidence base that examines agent capability,
        public software creation, and market response, distinguishing growing
        capacity to build from evidence of changed purchasing (\cref{sec:data}).
  \item We examine open-source alternatives to selected commercial products and a
        public case of rapid reimplementation with coding agents, distinguishing
        software creation from demonstrated reliability and adoption
        (\cref{sec:results}).
  \item We develop conditional capability projections and propose observable market
        tests to assess whether greater capacity to build translates into reduced
        software purchases (\cref{sec:scenarios}).
\end{enumerate}

\clearpage
\begingroup
\setlength{\intextsep}{5pt}
\datafigure[H]{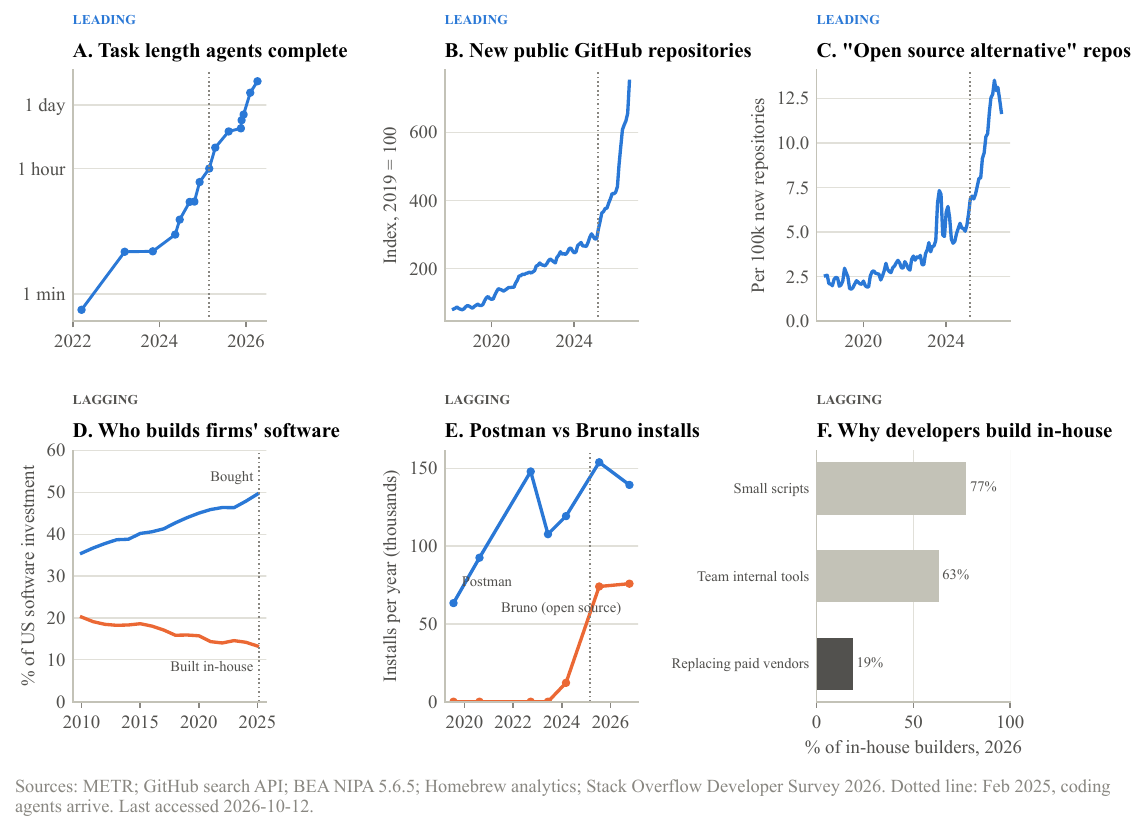}{\linewidth}
  {Leading indicators (top) and lagging indicators (bottom).}
  {fig:chain}

\begin{figure}[H]
\centering
\includegraphics[width=\linewidth]{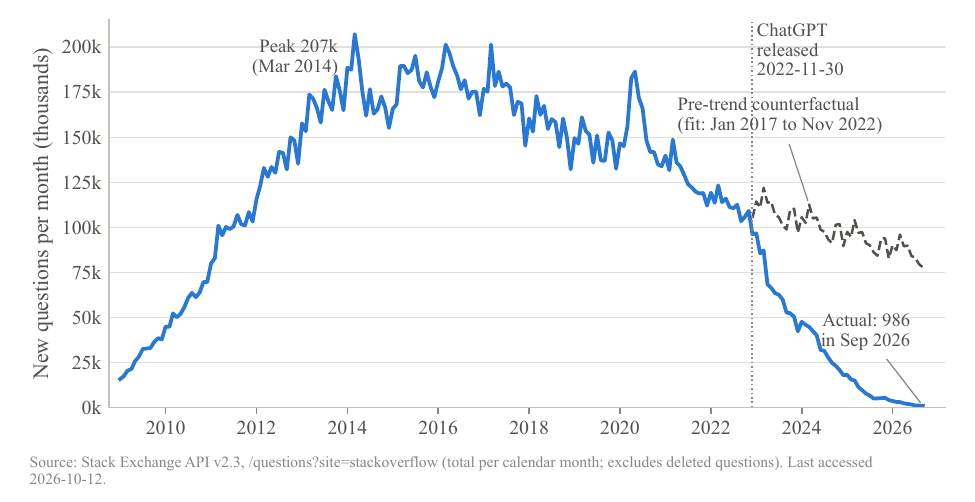}
\caption{New Stack Overflow questions per month.}
\label{fig:so}
\end{figure}
\endgroup
\clearpage

\section{Background and Related Work}
\label{sec:background}

\subsection{Make or buy, and the economics of software}
The make-or-buy decision compares the cost of organizing work inside a firm with the
cost of obtaining it from a supplier \citep{coase1937nature,williamson1975markets}.
Buying software has often been attractive because a vendor can spread development
costs across many customers. Digital products are cheap to reproduce
\citep{goldfarb2019digital}. A customer building for its own use cannot spread the
initial cost in the same way. A subscription can also bring updates, support and an
existing product that works with other systems. If agents make a working alternative
cheaper to create, those other sources of value still matter. As
\citet{teece1986profiting} explains, complementary assets such as distribution and
service can sustain a firm even when imitation becomes easier.

Open-source software is a partial precedent. Its development reflects motives beyond
direct sales \citep{lerner2002simple}; it can raise the productivity of firms that use
it \citep{nagle2019open}; and its economic replacement value is large
\citep{hoffmann2024value}. Yet available code is not a finished answer for every
customer. Someone must choose it, adapt it, verify it and keep it working. The question is whether agents make those steps cheap enough for open-source and customer-built
tools to replace narrower, paid applications.

\subsection{Building with AI}
\citet{sarkar2023code} anticipated that generative models could help more people write
software for themselves. Later conceptual work describes falling implementation
costs and a shift toward agentic software
\citep{dinu2026commoditization,cao2026agentic}. The distinction relevant to this
paper is what happens after a user asks for code. Chat assistants offer advice and
suggestions; a coding agent can inspect a project, edit files, run tests and revise its
work across steps. That may reduce the effort
needed to turn a stated need into a working application. It does not remove the need
to decide what the application should do or to check its behavior. The arrival of
coding agents is therefore a plausible point to examine in a time series, not a
causal experiment by itself.

\citet{klotz2026buyorbuild} argues that lower development costs are most
likely to shift build-or-buy decisions for commodity utilities and custom
applications that give organizations a competitive advantage. Regulated
and mission-critical systems remain harder to replace.
\citet{klein2026headless} model how agents could also lower the cost of coordinating
specialized software components. These are conditional arguments, not observations
that customers have left vendors. They identify the factors our evidence must
separate: initial construction, integration, verification and ongoing support.

\subsection{Evidence on AI and software work}
Studies of coding tools measure different stages of production. A controlled study of
an AI assistant found faster task completion \citep{peng2023impact}, while a randomized
trial with experienced open-source developers found that the tools available to them
slowed work in familiar repositories \citep{becker2025measuring}. Task gains cannot
be assumed to apply to every developer or codebase.

The closest empirical predecessor follows what happens after code is written.
\citet{demirer2026writing} examine how developers' use of AI coding tools
relates to commits, projects and releases. The reported gains are smaller
for projects and releases than for commits. They also find more new applications across software marketplaces without a corresponding
rise in total use. Their result shows why a surge in code or app creation may coexist
with little change in what customers use. Our study follows alternatives aimed at
named commercial products and public measures of market response. Neither study
directly measures customers canceling subscriptions after building a replacement.

Other work shows why output may stall between a repository and a dependable product.
\citet{song2024impact} find more contributions but also more coordination in
open-source projects using an AI assistant. \citet{he2026speed} find that an initial
gain in development speed came with a lasting rise in code complexity. Maintainers
merge many agent-assisted pull requests \citep{watanabe2025agentic}, but a merge does not prove an entire product is reliable. Software benchmarks document
improving task capability \citep{jimenez2024swebench,kwa2025measuring}, while
SaaSBench finds that configuring and integrating multi-component enterprise systems
remains difficult \citep{saasbench2026}. These studies make verification and upkeep
part of the economic question, not an afterthought.

Scientific software illustrates the role of domain knowledge. The CLAX-PT case
study reports a physicist supervising an AI coding agent and identifying an
invalid physical calculation despite passing numerical tests
\citep{nguyen2026physics}. It shows how a domain expert can help direct and evaluate
implementation, while also documenting failures that required specialist judgment.
The case concerns software production and validation; it does not measure a
commercial purchase displaced by the resulting tool.

\subsection{Evidence on AI and markets}
Adjacent markets offer early evidence of changing demand, but they measure different
outcomes. Stack Overflow activity was declining before ChatGPT;
\citet{delriochanona2024large} nevertheless find a further drop after its release
relative to comparison forums. This is evidence about demand for shared answers, not
about software subscriptions. \citet{demirci2025who} find weaker demand for freelance
work that is easier to automate relative to work requiring manual skills. Such tasks
may disappear, change or move to other suppliers without a customer building a
replacement application. Posting data also show a weaker market for junior software
roles \citep{westby2026entry}, though job advertisements respond to broader hiring
conditions as well as AI.

Building with AI may also strain the supply of reusable code. In a theoretical model,
\citet{koren2026vibe} predict that AI-mediated use of open-source code can reduce the
feedback and returns that support its maintainers. More easily created alternatives
would then draw on a shared resource whose quality could weaken. This is why we should measure continued use and maintenance, rather than count new projects alone.

\subsection{The gap}
These studies observe different links in the same chain. Capability studies ask
what agents can complete. Productivity studies follow code into releases. Platform
studies measure demand for answers or paid tasks. Economic theory explains why a
customer might build or buy. Few studies observe whether a customer stops purchasing
an application after gaining the ability to build an acceptable substitute. We bring
public measures of the successive links together over time and report evidence that
cuts against substitution as well as evidence that supports it. Direct customer-level
replacement and its effect on vendor revenue remain the decisive tests.

\section{Framework: The Software Substitution Frontier}
\label{sec:framework}

The unit of comparison is a customer's workflow. An organization using a
professional image editor may need only cropping, background removal and file
export. A tool that handles those tasks could replace the editor without matching every feature. The comparison includes the functions the customer needs, such as support for teamwork and connections to other systems. We
compare building and buying for a customer $c$ and workflow $w$; the same product
may be essential to one customer and replaceable for another.

Software creation, a viable replacement and paid substitution are distinct outcomes.
A script can count as a build if it performs a defined task. To become a viable
replacement, it must meet the customer's acceptance standard and total-cost test
over the intended period. Paid substitution additionally requires a reduced
purchase, a cancellation or a documented decision to build instead of buying. A
public alternative supplies a possible input to this choice; its creator need not
be a customer of the product it targets.

\Needspace{8\baselineskip}
The comparison follows established make-or-buy logic
\citep{coase1937nature,williamson1975markets}. Recent work applies it to agent-built
software and its lifecycle costs \citep{klotz2026buyorbuild}. We define the following
expressions to list the costs a direct test would need to measure. They are our
formulation, not estimates. Over $T$ months, the cost of buying is
\begin{equation}
  C^{\text{buy}}_{c,w}(T) = \sum_{t=1}^{T} p_{w,t} + I_{c,w} + A_{c,w} + S_{c,w},
\end{equation}
where $p_{w,t}$ is the price, $I$ is the cost of integration, $A$ is administration and
$S$ is the cost of being locked in. Building or adapting software incurs different
costs:
\begin{equation}
  C^{\text{build}}_{c,w}(T) = B_{c,w}(\theta) + V_{c,w}(\theta) + M_{c,w}
  + \sum_{t=1}^{T}\bigl(h_{c,w,t} + m_{c,w,t}(\theta)\bigr)
  + \lambda_{c}\,F_{c,w}(\theta),
\end{equation}
where $B$ is the cost to implement, $V$ to verify, $M$ to migrate, $h$ is hosting and
any model usage during operation, $m$ is maintenance, and $\lambda F$ is the expected
cost of failure. Model charges incurred while writing and testing belong in $B$
and $V$. Software written with AI need not use AI when it runs, so ongoing model
charges apply only when the tool requires them. Greater
agent capability $\theta$ may lower implementation cost. Whether it also lowers
verification, maintenance and failure costs is an empirical question. The customer
indices matter: a team that understands a workflow, controls its data and can check
outputs may face different build and verification costs from a buyer without those
resources. Domain knowledge can help specify requirements and detect errors, while
software engineering remains necessary to connect and operate the resulting system.

We call the set of workflows for which building is a credible alternative to buying
the substitution frontier. In our formulation, workflow $w$ lies inside it for
customer $c$ when
$C^{\text{build}}_{c,w}(T) + \Delta_{c,w} < C^{\text{buy}}_{c,w}(T)$, where $\Delta$
is the net value of factors that are hard to price, such as convenience, control and
trust. This adjustment can favor either choice. If agents lower total build cost
while the alternative remains acceptable, the frontier moves outward. Narrow,
self-contained workflows, such as recording a screen or merging PDF files, may cross
it before large, regulated or deeply integrated systems. Our question is whether
this frontier is now moving fast enough to observe.

A direct test of the frontier would specify a workflow and its acceptance standard
before an agent starts building. It would record model charges, human time,
verification, hosting, integration and later maintenance. An alternative would count
as viable only if it meets the standard and costs less than buying for the same
customer and period. Repeating this test across workflows and over time would show
whether the frontier is moving. This paper measures early traces of that movement;
it does not claim to have completed the full test.

Until those direct tests are available, the framework separates early signs of
building from later signs of market change. The early indicators are agent capability,
building cost, software creation and alternatives to commercial products. The later
indicators are installs, spending, survey answers about replacing vendors and vendor
revenue. A gap between the groups could precede a market shift. It could also persist
if maintenance, security, integration or trust prevent switching. \Cref{tab:hypotheses}
sets out these proposed links and the evidence that would count against each one.
They guide interpretation of separate indicators, not causal tests.

\begin{table}[H]
\centering
\small
\renewcommand{\arraystretch}{1.15}
\caption{Proposed links and evidence against them.}
\label{tab:hypotheses}
\begin{tabularx}{\linewidth}{@{}>{\raggedright\arraybackslash}X >{\raggedright\arraybackslash}X >{\raggedright\arraybackslash}X@{}}
\toprule
Proposed link & Measure & Counts against \\
\midrule
Agent capability on software tasks rises & Task time horizon; coding benchmarks; cost per score & Doubling time lengthens \\
People build more software & Repositories, packages, pushes, Show HN, AI-credited commits & No change beyond the prior trend \\
Alternatives to commercial products grow faster than software in general & Alternative repositories per 100k; product-level discovery & Normalized rate stays flat \\
Alternatives attract use & Installs, downloads, outside contributors & Use stays negligible \\
Buyers shift from buying to building & Software investment by type; surveys; commercial installs & Shares and installs stay unchanged \\
\bottomrule
\end{tabularx}
\end{table}

\section{Data and Methods}
\label{sec:data}
\label{sec:methods}

\subsection{Design}
This is an observational study of the links between agent capability, software
creation, use and market response. Each source measures a different link.
Repositories show what was created, installs show acquisition, and company results
reflect many kinds of demand. We compare these measures over time and report patterns
that disagree as well as those that agree. The observations concern models,
projects, platforms and aggregate markets rather than a panel of identified
organizational buyers. The activity and market series do not link an internal build
to a reduced purchase or cancellation by the same customer. An external organizational
survey adds self-reported decisions to build instead of buying, but no verified
purchase histories. The analysis can
therefore test parts of the proposed mechanism, but cannot measure a substitution
rate or identify the effect of agents. \Cref{tab:sources} maps each source to its role.
The code, notebooks, and data supporting these analyses are archived on Zenodo
\citep{malo2026researchdata}.

For the freelance marketplaces, we compare each firm's reported operating
measures with the same quarter a year earlier. We also compare changes in
their monthly adjusted share prices with SPY, a fund that tracks the
S\&P 500. These measures provide context on demand for freelance services
and investors' expectations. However, the platforms offer many services
beyond software development, and share prices reflect factors beyond AI.
These data therefore cannot establish whether customers are replacing
paid software with tools they build themselves.

The GitHub and Hacker News series describe public developer communities, and the
Stack Overflow survey respondents are self-selected developers. These populations help us examine
early building activity; they do not represent all organizational buyers or general
computer users. Published scientific and company accounts provide qualitative
context \citep{nguyen2026physics,airfocus2026archie}. We use them as selected
illustrations of supervision and build-or-buy choices, retain their reported
difficulties, and exclude them from the quantitative samples and estimated trends.

We also record McKinsey's organizational survey and Vals AI's application-building
benchmark as separate sources. The survey gathered \NumBuildSurveyN{} responses
across \NumBuildSurveyNations{} nations from \NumBuildSurveyFieldStart{} to
\NumBuildSurveyFieldEnd{}, with responses weighted by national contributions to global
GDP \citep{customer_building_mckinsey2026}. We report its purchasing question without
pooling it with the developer survey. For Vibe Code Bench, we archive the leaderboard
and select \NumVcbSelectedN{} recent models on the common OpenHands harness, including
the leading entry and models from several providers. Models build web applications
from short specifications in a prepared environment, with up to \NumVcbLimitHours{}
hours or \NumVcbLimitTurns{} turns per build. An application earns credit
for a UI test when at least \NumVcbSubstepPassPct{} percent of its steps succeed;
the overall score averages those test results across applications
\citep{customer_building_vals2026,tran2026vibecode}. These scores measure tested
functions, not complete applications or METR task hours.

\begin{table}[H]
\centering
\small
\renewcommand{\arraystretch}{1.1}
\caption{Sources and measures.}
\label{tab:sources}
\begin{tabularx}{\linewidth}{@{}l >{\raggedright\arraybackslash}X >{\raggedright\arraybackslash}X@{}}
\toprule
Link & Public source & Measure \\
\midrule
Capability & METR horizons \citep{agent_capability_cost_metr_th11}; Epoch ECI and FrontierSWE \citep{agent_capability_cost_epoch_eci,agent_capability_cost_epoch_frontierswe}; Vibe Code Bench \citep{customer_building_vals2026}; Aider \citep{agent_capability_cost_aider_leaderboard} & Task hours, separate capability and functional test scores, and run cost \\
Creation & GitHub search, commit search and Innovation Graph \citep{github_repo_trends_searchapi,ai_attribution_commits_searchapi,innovation_graph_dataset}; package registries; Hacker News \citep{hacker_news_algolia_api} & Repositories, pushes, packages, posts and AI-credited commits \\
Alternatives & GitHub repositories, product sites, curated lists and Show HN posts & Candidates for a fixed commercial-product sample \\
Use and attention & Homebrew \citep{homebrew_analytics_2026}; Wikipedia pageviews \citep{wikimedia_pageviews_api_2026}; GitHub releases and issues & Installs, attention, downloads and outside engagement \\
Buying and work & BEA \citep{bea_software_investment_table565}; Stack Overflow and McKinsey surveys \citep{stackoverflow_survey2026_buildtool,customer_building_mckinsey2026}; Indeed and BLS \citep{labor_market_indeed_softdev,labor_market_bls_ces} & Software investment, reported building and purchasing decisions, postings and employment \\
Adjacent markets & Upwork and Fiverr quarterly reports \citep{freelance_upwork_q2_2026,freelance_fiverr_q2_2026}; adjusted share prices \citep{freelance_yahoo_prices} & Marketplace activity, clients, revenue and valuation context \\
\bottomrule
\end{tabularx}
\end{table}

\subsection{Product sample and alternatives}
We fixed a sample of \NumTenProducts{} commercial products before searching for
alternatives: screen recording (Screen Studio, Loom), writing (Grammarly, QuillBot), PDF
tools (Adobe Acrobat, Smallpdf), image and design (Adobe Photoshop, Canva), productivity
(Notion, Trello) and developer tools (Postman, Retool). Choosing products first avoids
picking only the cases where alternatives went viral.

We used two definitions of an alternative. The narrow rule requires a repository
description to name the commercial product as an alternative, and the repository must
be active, not a fork, and have at least \NumAltNarrowMinStars{} stars. This rule misses
projects that describe only their function. The broad search adds candidates from web lists,
GitHub searches by product and workflow, GitHub topics, curated open-source lists
and Show HN posts. Open-source candidates enter the broad count if screening finds a match to the
product's core workflow, are neither archived nor forks, and have at least
\NumAltMinStars{} stars. Paid and closed replacements are recorded separately. The
first author's own repository is excluded from both searches. For some products, workflow
matching relies partly on keyword rules, so the broad count is an upper bound until
independent validation.

\subsection{Measuring change over time}
We also screen package creation counts to reduce short-lived entries. A new PyPI
project qualifies if it has a second release within \NumPkgFilterWindowDays{} days.
A package on crates.io qualifies if it has more than one published version
when the data are collected. A RubyGems package qualifies if it remains
listed in the registry. We omit recent PyPI and crates.io cohorts whose follow-up
window is incomplete. Packagist supplies only net additions from its aggregate series.
These screens can also exclude legitimate projects with a single release.

We use segmented regression for package creation and download series, where enough
monthly observations are available on both sides of the events. The model permits a
change in level and trend after ChatGPT's release, then another after the initial
coding-agent preview:
\begin{equation}
\begin{aligned}
  \log y_t ={}& \beta_0 + \beta_1 t + \beta_2 D^{\text{GPT}}_t
    + \beta_3 (t - t_1) D^{\text{GPT}}_t \\
    &+ \beta_4 D^{\text{agent}}_t + \beta_5 (t - t_2) D^{\text{agent}}_t
    + \gamma^{\top} s_t + \varepsilon_t,
\end{aligned}
\end{equation}
Here $y_t$ is a positive monthly count, $t_1$ is \NumPkgChatgptBreak{}, and $t_2$ is
\NumPkgAgentBreak{}. Each indicator $D_t$ switches on at its break. Its coefficient
captures an immediate level change; the interaction with time captures a change in
slope. The vector $s_t$ accounts for month-of-year patterns. We use
heteroskedasticity and autocorrelation consistent standard errors with a full year
of monthly lags \citep{newey1987simple}. \Cref{tab:its-packages} reports the fits.
As a sensitivity check, we move the coding-agent break across the six months on
either side of the chosen date while holding the model and sample fixed.
\Cref{tab:break-sensitivity} reports how the estimated slope change varies. This
check does not identify the date or cause of a change.

The Stack Overflow question series has a separate single-break fit and a pre-release
trend counterfactual. Other time series are described rather than fitted with this
model. Where a suitable denominator exists, we also show rates: alternative
repositories per new repository, AI-credited commits per public commit, and
alternative Show HN posts per Show HN post. Those rates help separate changes in a
specific activity from growth of its platform. Break dates mark events, not random
assignments. The coefficients and counterfactuals are descriptive associations.

\subsection{Capability projection}
METR's task horizon measures task length by how long a human expert would
take to complete the work. It estimates the task length at which an agent
achieves a specified success rate. It is not the time an agent can work unattended. We fit a
straight line to the logarithm of the \NumProjSuccessFifty{} and
\NumProjSuccessEighty{} percent horizons of frontier models against release date. The
recent fit starts in \NumProjFastStartYear{}; the longer fit starts in
\NumProjSlowStartYear{}. Following the source's fitting rule, we omit models whose
\NumProjSuccessFifty{} percent horizon exceeds \NumProjMaxFitHours{} hours.

We extend each fitted line to the point at which its task horizon equals a skilled
human workday (\NumProjWorkdayHours{} hours), workweek
(\NumProjWorkweekHours{} hours), workmonth (\NumProjWorkmonthHours{} hours) or workyear
(\NumProjWorkyearHours{} hours). We resample models \NumProjBootstrapSamples{} times
and report the \NumProjBandLowPercentile{}th to \NumProjBandHighPercentile{}th
percentiles of the resulting paths and crossing dates. These bands reflect variation
across the selected models, not uncertainty in METR's task sample. The dates are an
extrapolation if the fitted trend continues, not a forecast of product replacement.

For a dated check, we fit the trend only to frontier models released before the
coding-agent break and compare its predictions with later measured models. This
tests how the earlier capability trend tracked later benchmark results. It cannot
validate the future path or tell us when buyers will leave vendors.

\subsection{Case study}
The Craft suite is an illustrative frontier case, selected because its public build
was unusually rapid and visible. It is not a representative sample. We collected
the \NumCraftApps{} application repositories from the suite's GitHub organization
\citep{craft_github_org} through the GitHub API. Commit histories use a fixed cutoff;
other changing measures are retrieval-time snapshots. The measures include default-branch
commits, explicit AI co-author trailers and author accounts, agent instruction files, automated workflow
runs, issues, pull requests and release-asset downloads. We counted source lines
from repository clones. Commit counts depend on merge practice, AI trailers miss
uncredited use, and downloads do not identify distinct users.

We compare these public measures with roadmaps and maturity claims on the suite's
website \citep{craft_site_apps}. The case documents speed, scale and outside engagement. It
does not establish feature parity, a clean-room development process or a customer
switching away from a paid product.

\section{Results}
\label{sec:results}

The evidence in \cref{fig:chain} shows growing capacity to build software
alongside mixed evidence of changes in purchasing. Agent capability has
improved over several years, while public software creation and the
appearance of commercial alternatives accelerated around the start of
the coding-agent era. Aggregate indicators show continued purchasing,
although some survey respondents report building instead of buying.
The available data do not establish how widespread this shift is.
We examine each link in turn, including evidence that challenges
the proposed shift from buying to building.

\subsection{Agents can do much more, for much less}
\label{sec:res-capability}
Frontier agents can now complete increasingly long software
tasks. Since \NumProjFastStartYear{}, their estimated task horizon at
\NumProjSuccessFifty{} percent success has doubled about every
\NumCapMetrFitDoublingDays{} days
(panel A of \cref{fig:metr}; \NumCapMetrFitCiPercent{} percent interval:
\NumCapMetrFitBootCiLow{} to \NumCapMetrFitBootCiHigh{} days).
This corresponds to roughly a \NumCapMetrFitGrowthPerYear{}-fold increase
in task horizon per year. GPT-4's estimated horizon was
\NumCapMetrGptFourPFifty{} of skilled human work. The largest reported
estimate, for a model released in \NumCapMetrTopDate{}, is
\NumCapMetrTopPFifty{}. However, this estimate exceeds METR's reliable
measurement range of \NumCapMetrReliableMaxHours{} hours. At
\NumProjSuccessEighty{} percent success, the same model's horizon is
\NumCapMetrTopPEighty{}. The series covers models released through
\NumCapMetrLastRelease{} and is no longer actively updated
\citep{agent_capability_cost_metr_th11}, so it does not establish
the task horizons of newer models.

Panel B of \cref{fig:metr} provides evidence on newer models.
Epoch's general capability index includes \NumCapEciRecentN{} models
released in \NumCapEciLatestMonth{}, including GPT-6 Sol, GPT-6 Astra
and Claude Opus 5.5 \citep{agent_capability_cost_epoch_eci}.
The index combines results from several benchmarks. It does not measure
task length in human working hours or the reliability of finished software.
Overlapping intervals also make it difficult to distinguish the
leading models.

FrontierSWE provides a separate comparison of coding performance,
evaluating recent Gemini, Grok and open-weight models under the same
testing setup (\cref{fig:app-swe}). The highest mean normalized task
score is \NumCapFrontierSweTopScore{}
\citep{agent_capability_cost_epoch_frontierswe,agent_capability_cost_frontierswe}.
The bars show the lowest and highest scores across
\NumCapFrontierSweTrials{} runs, rather than statistical uncertainty
intervals. These results extend coverage to newer models, but do not
establish whether an agent can build and maintain a complete product.

Vibe Code Bench tests a closer step toward customer building: creating a web
application from a short specification. \Cref{fig:vibe-code} shows selected recent
models on the same OpenHands harness. Their scores range from
\NumVcbLowSelectedScore{} to \NumVcbTopScore{} percent in the archived leaderboard
\citep{customer_building_vals2026}. These are averages of functional UI-test
results, not the share of applications that are fully correct. The prepared
environment also removes some setup work a customer would face. The results show
that recent models can implement much of the tested functionality, while leaving
security, integration and continued operation unmeasured
\citep{tran2026vibecode}. Together with the other recent benchmarks, this evidence
extends model coverage without supplying new observations for METR's time-horizon
curve.

\datafigure[H]{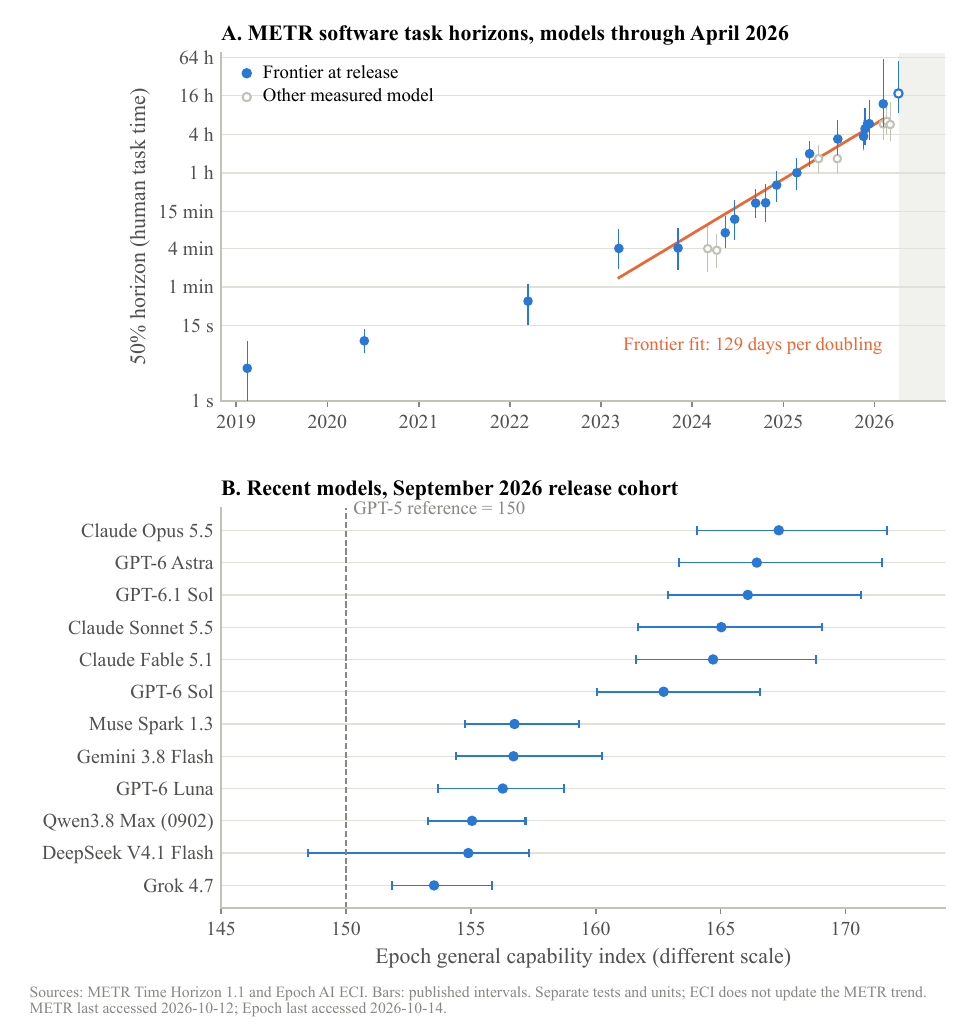}{\linewidth}
  {Historical task horizons and recent model capability.}
  {fig:metr}

\datafigure[H]{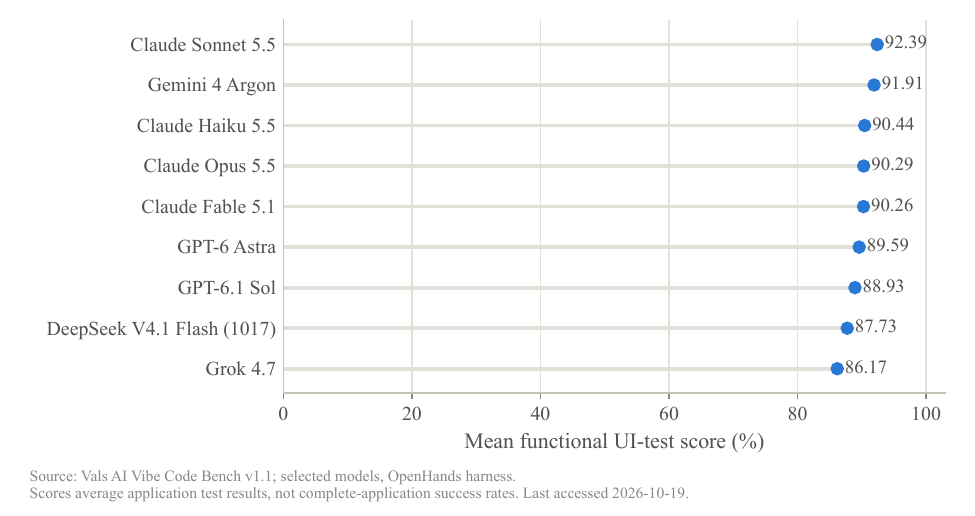}{\linewidth}
  {Functional test scores for selected application-building agents.}
  {fig:vibe-code}

The cost of a fixed benchmark result also fell (\cref{fig:aider-cost}). On the Aider coding benchmark, the cheapest
run that solved at least 60 percent of tasks cost \$\NumCapAiderCostSixtyFirst{} in
\NumCapAiderCostSixtyFirstMonth{} and \$\NumCapAiderCostSixtyLast{} by
\NumCapAiderCostSixtyLastMonth{}, a drop of \NumCapAiderCostSixtyFold{} times in eight
months. This is the cost of a benchmark run, not a
deployed and maintained product.

\datafigure[H]{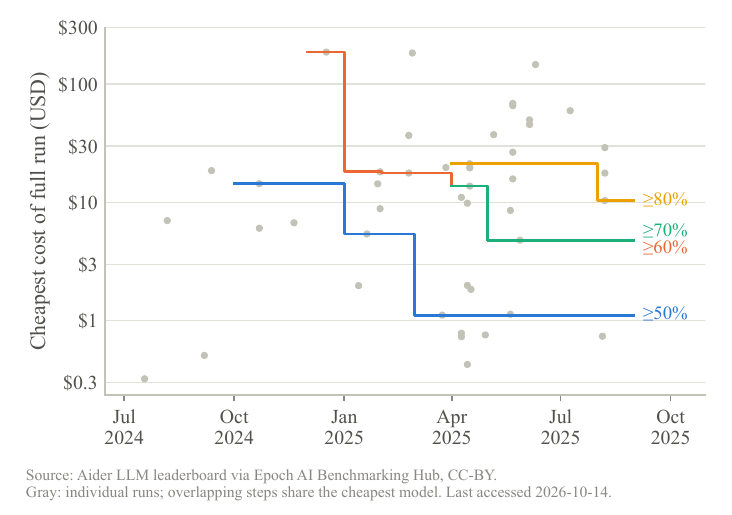}{0.75\linewidth}
  {Cost of reaching a fixed score on the Aider benchmark.}
  {fig:aider-cost}

\subsection{Public software creation accelerates in the agent era}
\label{sec:res-supply}
If agents lower the cost of building, people may build more. Public creation measures
rise, and their timing is telling. \Cref{fig:packages} shows new packages per month on four registries. During
the ChatGPT period, from December 2022 to January 2025, growth in new packages slowed. On
PyPI, average yearly growth fell from \NumPkgPypiYoYPre{} percent before ChatGPT to
\NumPkgPypiYoYChatEra{} percent. On crates.io it fell from \NumPkgCratesYoYPre{} to
\NumPkgCratesYoYChatEra{} percent. After agents arrived in February 2025, growth jumped to
\NumPkgPypiYoYAgentEra{} percent on PyPI and \NumPkgCratesYoYAgentEra{} percent on
crates.io. In the time series model, the trend in screened new crates rose by
\NumPkgCratesFilteredAgentSlope{} annualized log points after February 2025 and the
trend in screened PyPI projects by \NumPkgPypiFilteredAgentSlope{} log points
(both $p<0.001$;
\cref{tab:its-summary}). The acceleration is consistent with faster supply
growth during the agent period. It is not specific to the February break:
the slope changes remain positive when that break is moved across nearby months,
but the best-fitting date differs by registry (\cref{tab:break-sensitivity}).
These time series cannot establish that agents caused the growth; registry
policies, spam and changes in the developer population may also matter.

\datafigure[H]{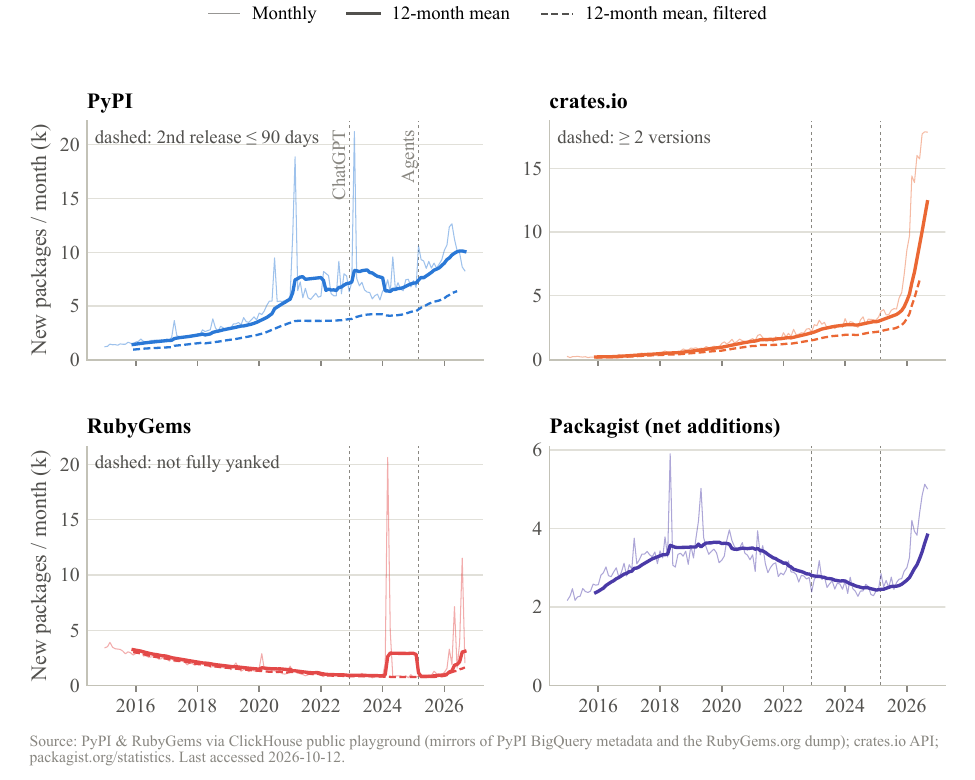}{\linewidth}
  {New packages per month on four registries.}
  {fig:packages}

\begin{table}[H]
\centering
\small
\caption{New-package growth after ChatGPT and coding agents.}
\label{tab:its-summary}

\begin{tabular}{@{}llrrr@{}}
\toprule
 & & \multicolumn{2}{c}{Annualized slope change (log points)} & \\
\cmidrule(lr){3-4}
Registry & Series & \shortstack{ChatGPT\\Dec 2022} & \shortstack{Coding agents\\Feb 2025} & Months \\
\midrule
PyPI & Screened projects & $-15.5^{***}\;(3.0)$ & $+17.9^{***}\;(3.8)$ & 126 \\
crates.io & Screened crates & $-19.3^{***}\;(2.4)$ & $+119.6^{***}\;(14.3)$ & 126 \\
RubyGems & Screened gems & $+10.4^{***}\;(1.8)$ & $+84.9^{***}\;(4.9)$ & 129 \\
Packagist & Net additions & $-5.8^{***}\;(2.2)$ & $+55.1^{***}\;(6.4)$ & 129 \\
\bottomrule
\end{tabular}

\par\smallskip
\parbox{\linewidth}{\footnotesize Estimates are changes in the slope of log monthly
counts, annualized and multiplied by 100. Standard errors are in parentheses. Each
model includes both dates and month-of-year effects, with standard errors adjusted
for autocorrelation. Screened series use the registry-specific rules in Methods.
Packagist reports net additions. $^{*}p<0.1$, $^{**}p<0.05$,
$^{***}p<0.01$. Unscreened estimates, level shifts and downloads appear in
\cref{tab:its-packages}.}
\end{table}

\Needspace{8\baselineskip}
GitHub data show that the rise in public development activity extends
beyond package creation (\cref{fig:ig-activity}). In the first quarter
of 2026, code pushes increased by \NumIgPushesYoY{} percent from a year
earlier, compared with an earlier annual average of
\NumIgPushesYoYPreAgentMean{} percent. The increase was geographically
broad: \NumIgPushEconomiesGrowingPct{} percent of the
\NumIgPushEconomiesN{} economies in the data recorded growth.
The developer account total increased by \NumIgDevNetAddLast{} million
that quarter, compared with an average quarterly increase of
\NumIgDevNetAddTwentyThreeTwentyFour{} million in 2023 and 2024.
New public repositories also increased by
\NumGhtRepoAllGrowthTwelveM{} percent over the preceding twelve-month
period (\cref{fig:app-repos}). Language participation changed as well
(\cref{fig:app-lang}), although these shifts began before coding agents.

Hacker News provides a separate signal of increased project sharing.
Show HN posts averaged \NumHnShowLastTwelveMean{} per month over the
last year, compared with \NumHnShowPreMean{} in early 2022. Their share
of all stories rose from \NumHnShowShareStoriesPre{} to
\NumHnShowShareStoriesLastTwelve{} percent (\cref{fig:app-showhn}).
These measures indicate greater public activity, but do not
establish how much additional usable software was produced or whether
coding agents caused the increase. Developer and repository totals
include inactive accounts and projects, pushes measure quarterly
activity, and Show HN posts measure sharing rather than verified
software completion.

\datafigure[H]{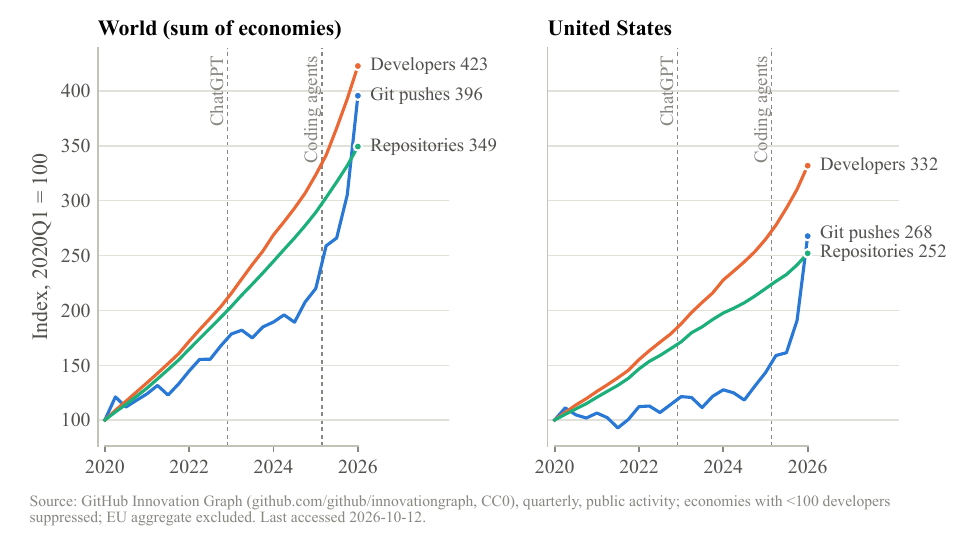}{\linewidth}
  {GitHub developers, repositories and pushes.}
  {fig:ig-activity}

Some new code is visibly linked to agents, and that signal is growing
(\cref{fig:aicommits}). In September 2026, the search counted
\NumAicTotalLastMonthM{} million public commits credited to an AI tool,
\NumAicTotalFoldYoY{} times the
search count a year earlier. Claude Code accounts for
\NumAicClaudeShareOfAILastMonth{} percent of the attributed commits. Many uses
leave no credit line, while overlapping signatures and mirrored repositories
can duplicate counts. About
\NumAicStockAgentsMd{} public
files named AGENTS.md and \NumAicStockClaudeMd{} named CLAUDE.md, which hold instructions
for coding agents, now exist on GitHub. Commit messages mentioning these files also
increased (\cref{fig:app-instr}). Those mentions record configuration work, not a
count of finished products. Optional attribution, incomplete search results
and changing commit habits limit what
these counts can say about the amount of finished software.

\datafigure[H]{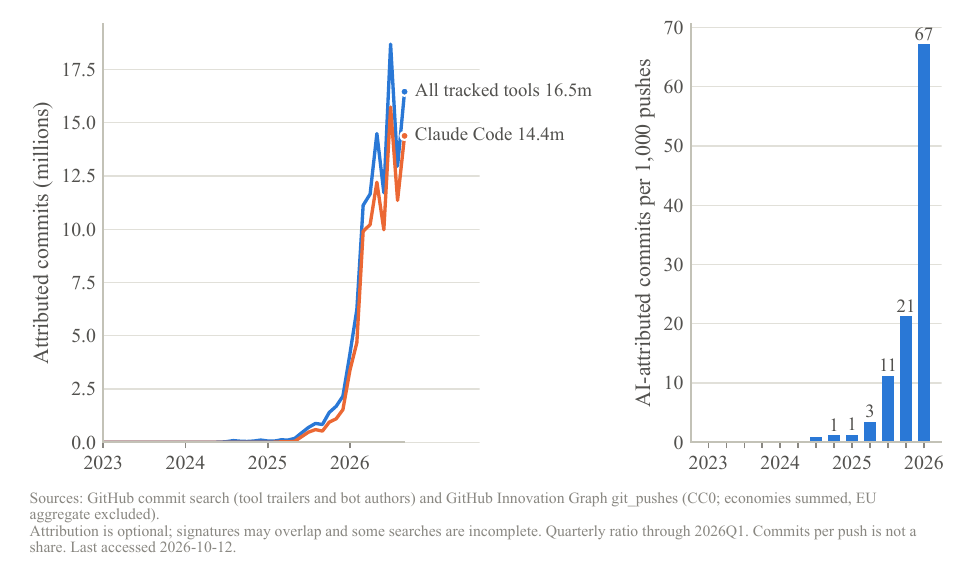}{\linewidth}
  {AI-attributed commits: monthly totals (left) and quarterly commits per push (right).}
  {fig:aicommits}

\subsection{More projects target commercial software}
\label{sec:res-alternatives}
Building more is not the same as building substitutes. To examine whether projects
increasingly target commercial functionality, we count repositories that describe themselves as an
alternative, relative to all new repositories (\cref{fig:altrate}). Repositories calling
themselves an ``open source alternative'' rose from \NumGhtOssAltRateTwentyTwo{} per
100,000 new repositories in 2022 to \NumGhtOssAltRateLastTwelveM{} in the last twelve
months. ``Self-hosted alternative'' rose from \NumGhtSelfhostAltRateTwentyTwo{} to
\NumGhtSelfhostAltRateLastTwelveM{}. Among repositories with at least ten stars, which
filters out most empty projects, the open-source alternative rate rose from
\NumGhtOssAltSTenRateTwentyTwo{} to \NumGhtOssAltSTenRateLastTwelveM{} per 100,000. The
increase is concentrated after February 2025.

\datafigure[!htb]{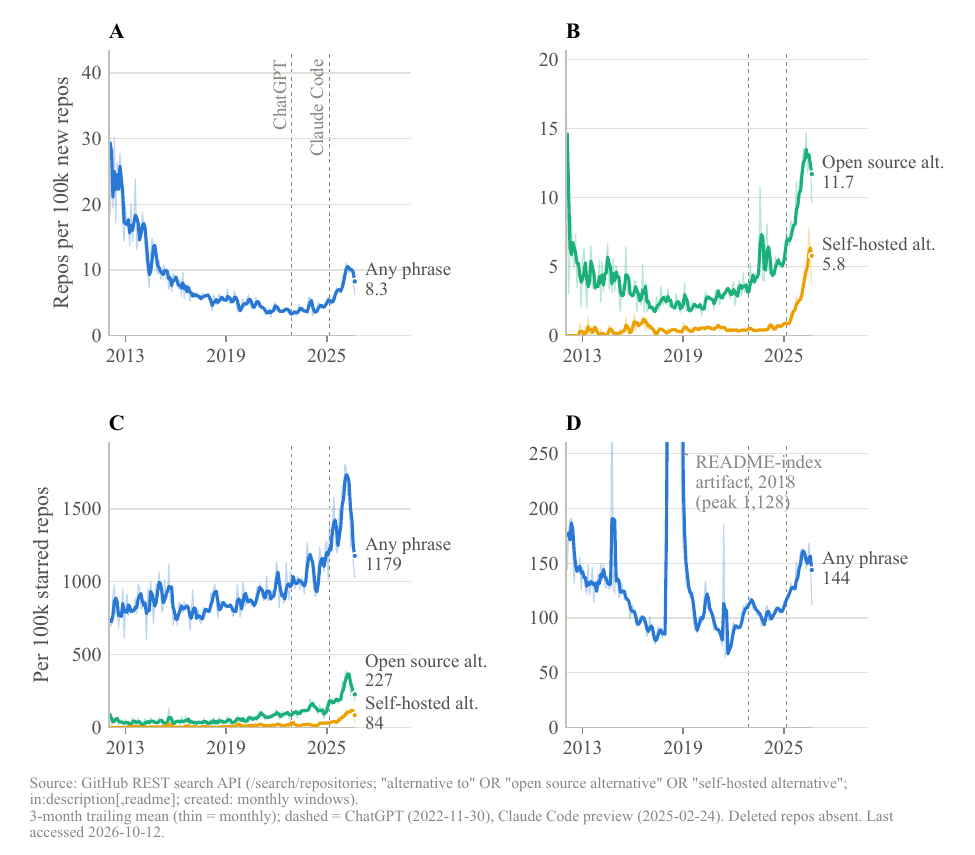}{\linewidth}
  {Alternative-repository rates: (A) descriptions only; (B) substitute phrases; (C) star-filtered matches; (D) descriptions or READMEs.}
  {fig:altrate}

At the individual-product level, the picture is similar, with an important
qualification (\cref{fig:broad}). For the
twelve products in our sample, the broad search found \NumAltCandidates{} candidates, of
which \NumAltIncluded{} open-source projects passed the automated screen, against only
\NumAltNarrowTotal{} under the narrow rule. The narrow timeline in
\cref{fig:app-narrow} shows that named alternatives existed well before coding agents,
while the broader search finds many more candidates. Workflow matches for some
products still need independent validation, so these screened counts are an
upper bound on qualified alternatives. Of the screened projects,
\NumAltIncludedAgentEra{} (\NumAltAgentEraPct{} percent) were created after February 2025.
The newest alternatives
also look different: \NumAltAiFilesPctAgentEra{} percent of those created after
February 2025 contain agent instruction files, against \NumAltAiFilesPctBefore{} percent
of older ones. The categories differ as the framework predicts. All
\NumAltScreenStudioBroad{} screened candidates for Screen Studio, a narrow screen
recording tool, appeared after February 2025, and \NumAltSmallpdfAgentEra{} of
\NumAltSmallpdfBroad{} for Smallpdf did. QuillBot, a paraphrasing tool, received no new
screened candidates. A general assistant could cover some paraphrasing tasks
without anyone publishing a separate tool, but this study does not measure that use.

\datafigure{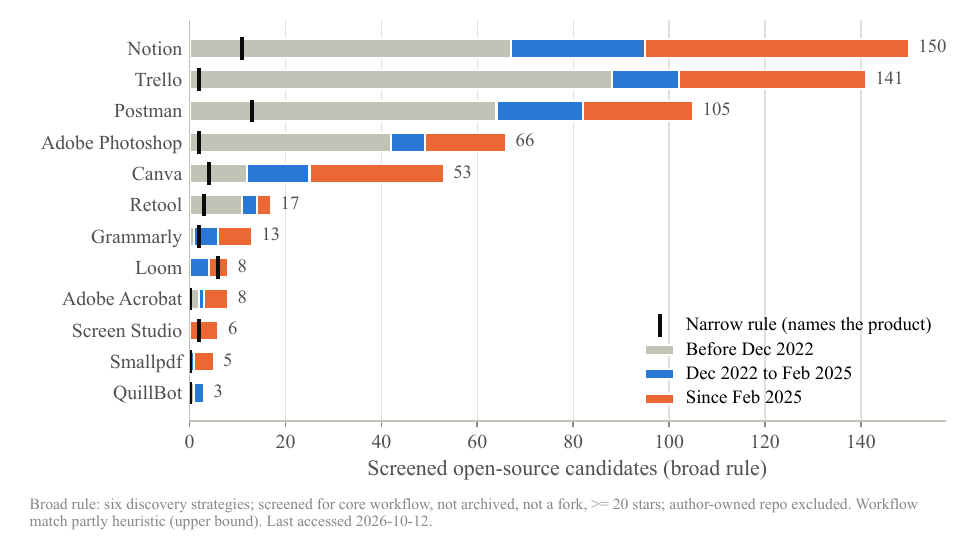}{\linewidth}
  {Screened open-source candidates for twelve products, by creation date.}
  {fig:broad}

\Needspace{8\baselineskip}
Not every measure points the same way. The share of Show HN posts that call themselves an
alternative fell slightly, from \NumHnAltShareTwentyTwo{} percent in 2022 to
\NumHnAltShareLastTwelve{} percent in the last year, so the new projects on Hacker News
grew in number, not in focus on substitution. The separate narrow pilot also raises a
question about durability: newer alternatives have a median of
\NumTenContribPost{} contributors, against \NumTenContribPre{} among those created before
ChatGPT \citep{github_rest_api_2026}. This is a small, selected sample, and older
projects have had more time to recruit. The contrast does not establish which
alternatives will last.

\subsection{The frontier case: a suite built at speed}
\label{sec:res-craft}
One visible example of what agents can now do is the Craft suite, an open-source
project that set out to reimplement major creative and office software
(\cref{fig:craft}). Between \NumCraftCreatedFirst{} and \NumCraftCreatedLast{}, its
authors created \NumCraftApps{} applications aimed at Adobe Photoshop, Lightroom,
Premiere Pro, Illustrator, After Effects, InDesign and Acrobat, Microsoft Word, Excel and
PowerPoint, Autodesk AutoCAD and Avid Pro Tools.

\datafigure{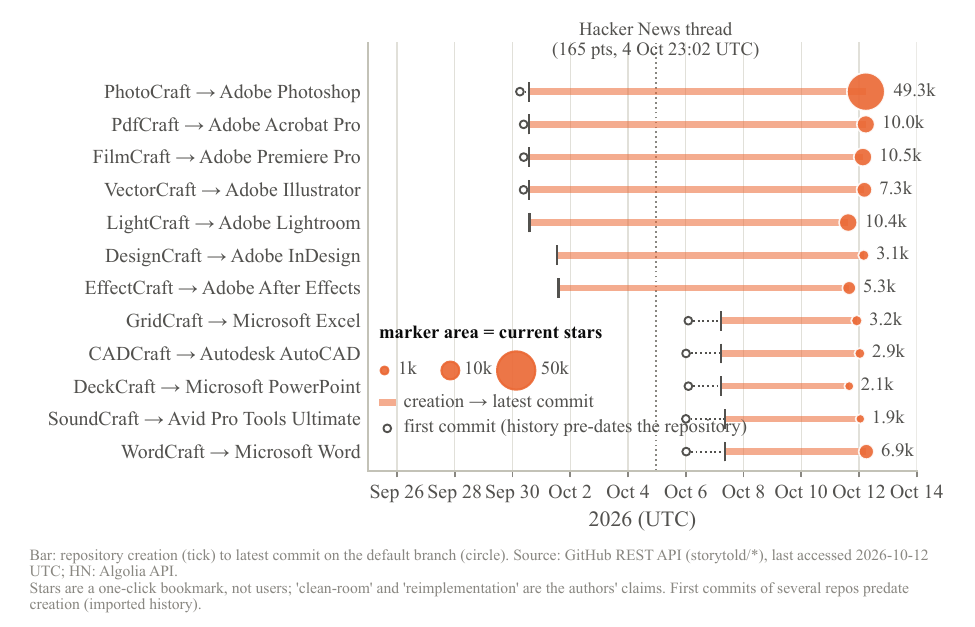}{\linewidth}
  {The Craft suite: twelve applications and the products they target.}
  {fig:craft}

The measured scale is large. In \NumCraftHistoryDays{} days of public history the suite
received \NumCraftCommitsTotal{} commits and grew to \NumCraftLocTotal{} lines of Rust,
with \NumCraftTestFns{} test functions. At least \NumCraftAiAnyPct{} percent of commits
carry an AI co-author line, most often Claude Opus 5.5
(\cref{tab:craft,fig:app-craftcommits}). Every application has an AGENTS.md file that
tells agents to work autonomously and to ask a human only about licensing, publishing
or secrets \citep{craft_github_org}. The project also drew outside engagement:
\NumCraftIssueReportersNonMember{} people outside the project filed issues,
\NumCraftPrAuthorsNonMember{} outside authors opened pull requests, and release files were
downloaded \NumCraftDownloadsTotal{} times. These are signs of engagement, not counts of
distinct users or canceled subscriptions. Downloads include automatic updates and
ancillary files.

The suite is not yet a replacement for the products it targets. Its own roadmaps label
nine applications alpha and three pre-alpha, and rate them a median of
\NumCraftReadyMedian{} percent ready for real work. Press coverage was mostly mixed or
skeptical, pointing to missing features, memory use, and open legal questions about the
term clean room. We treat the suite as a measure of the frontier, not of the market. It
shows how quickly agents and many contributors can produce a large codebase. It does not
show how quickly they can produce a dependable commercial substitute. The lagging indicators answer whether customers come to rely on such software.

The app-level evidence in \cref{tab:craft} makes the trade-off visible. Some apps have
large codebases and many contributors, yet the shares of successful automated runs
vary. Those runs include pull requests, and the roadmaps still rate readiness well
below full use. The daily record in
\cref{fig:app-craftcommits} shows bursts of AI-attributed and other commits. It records
output and participation, not independent evidence that the apps meet their target
products' quality or feature coverage. This is a collaborative build, not a measured
case of one customer canceling a subscription.

\subsection{Purchasing evidence is mixed}
\label{sec:res-lagging}
The lagging indicators tell a different story. In US national accounts
(\cref{fig:bea}), software that firms build for themselves fell from
\NumBeaShareOwnAccountTwentyTwentyTwo{} percent of software investment in 2022 to
\NumBeaShareOwnAccountLast{} percent in 2025, while purchased packaged software rose from
\NumBeaSharePrepackagedTwentyTwentyTwo{} to \NumBeaSharePrepackagedLast{} percent. This
measure has limits. BEA values in-house software by what firms pay their developers, so
if agents let the same staff build more, the series will not show it. Still, there is no
sign yet of a national shift from buying to building.

\datafigure[!htbp]{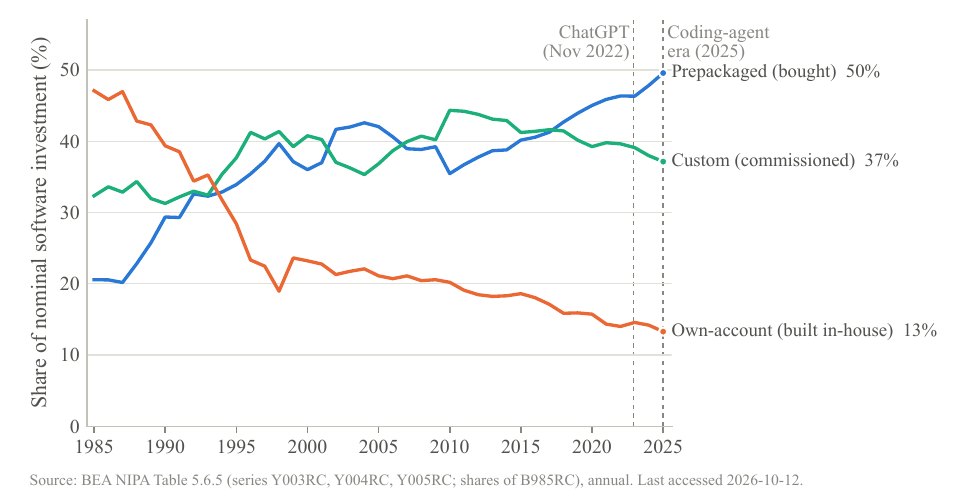}{0.9\linewidth}
  {US private software investment by type.}
  {fig:bea}

The Homebrew pilot has not shown a general shift away from commercial tools
(\cref{fig:homebrew}). Yearly installs of
Postman rose from \NumTenBrewPostmanFirst{} in 2019 to \NumTenBrewPostman{} in 2026, and
installs of Notion from \NumTenBrewNotionFirst{} to \NumTenBrewNotion{}. The open-source
alternative AppFlowy reached only \NumTenBrewAppFlowyPctNotion{} percent of Notion's
installs. The one strong exception is Bruno, an open-source alternative to Postman, which
grew from almost zero to \NumTenBrewBruno{} installs, or \NumTenBrewBrunoPctPostman{}
percent of Postman's. Bruno was created before ChatGPT. These data do not show
whether agents contributed to its later development or adoption.

These public measures miss another possible route to substitution. A customer could
use a general AI subscription for a narrow task or ask an agent to build a private
tool for it. Neither choice needs to produce a public repository or a Homebrew install.
The public activity series cannot establish how often either choice replaces a
paid product. Surveys can ask about the decision more directly.

\datafigure[!htbp]{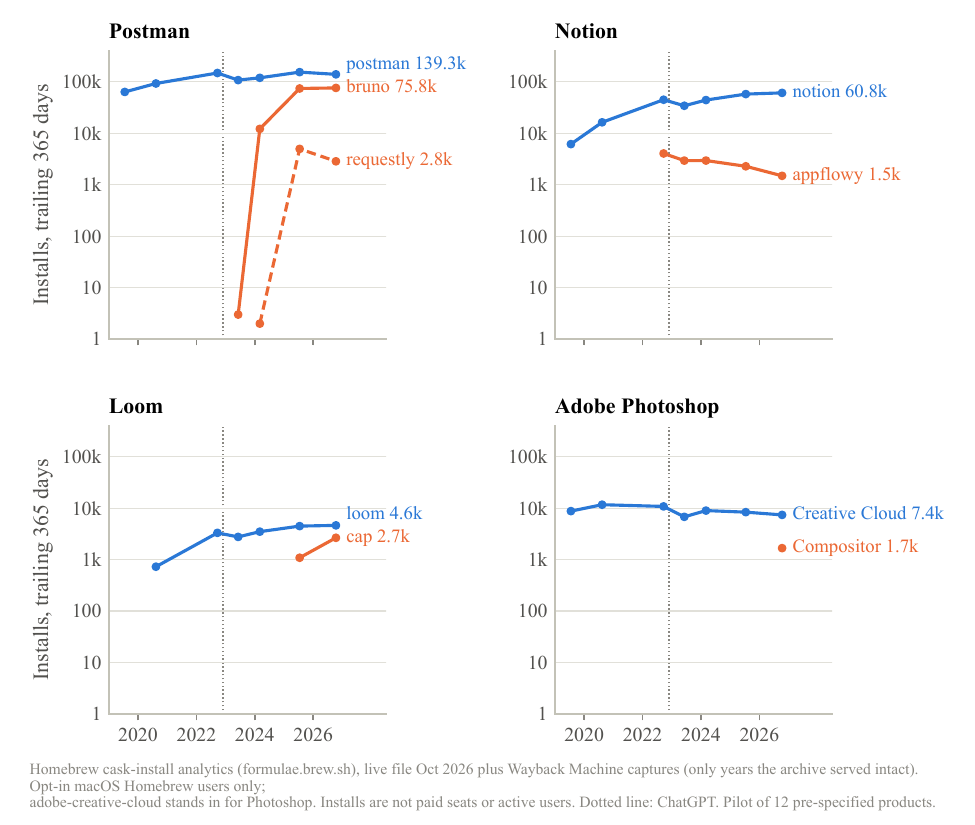}{0.95\linewidth}
  {Yearly installs of commercial tools and open-source alternatives.}
  {fig:homebrew}

The developer survey records the functions of in-house software. In the 2026 Stack
Overflow survey, \NumSoSurveyBuildN{} respondents who build in-house reported their
workflows. Most build small scripts
(\NumSoSurveySmallScripts{} percent) or internal tools for their team
(\NumSoSurveyInternalTools{} percent). Only \NumSoSurveyReplacingVendors{} percent said they
were replacing paid vendors. These are self-reported functions, not independently
verified cancellations or a representative survey of software-buying organizations.

McKinsey's organizational survey asks a more direct purchasing question. It reports
that \NumBuildSurveyForegonePct{} percent of respondents said their organization
had decided against buying at least one software product or feature because it
could build the functionality internally with coding agents
\citep{customer_building_mckinsey2026}. This reflects reported purchases
forgone, which can occur without a canceled subscription. It does not measure
spending displaced, continuing use or the quality of the internal tools. The
survey and developer responses concern different populations and questions, so
their percentages should not be compared as a trend. They suggest that some buying
decisions are changing while leaving the aggregate scale unresolved.

Employment at software publishers is also steady, at
\NumLabEmpSoftwarePublishersLast{} thousand in \NumLabCesLastMonth{}, close to its peak of
\NumLabEmpSoftwarePublishersPeak{} thousand in \NumLabEmpSoftwarePublishersPeakDate{}. Job postings for software
developers fell sharply (\cref{fig:app-jobs}), from an index of \NumLabSoftDevPeak{} in \NumLabSoftDevPeakDate{}
to \NumLabSoftDevAtChatGpt{} when ChatGPT was released and \NumLabSoftDevLast{} today, but
most of that fall came before ChatGPT, and postings have risen
\NumLabSoftDevSinceJanTwentyFivePct{} percent since January 2025.

Freelance marketplaces give an early, adjacent market signal. In their latest reported
quarter, Fiverr's annual active buyers fell \NumFreeFiverrBuyersDeclinePct{} percent and
its marketplace revenue fell \NumFreeFiverrMarketplaceRevenueDeclinePct{} percent from
a year earlier. Its management cited AI-related demand and traffic headwinds in work
most exposed to automation, while spend per buyer rose
\citep{freelance_fiverr_q2_2026}. Upwork's active clients
fell \NumFreeUpworkClientsDeclinePct{} percent and gross services volume fell
\NumFreeUpworkGsvDeclinePct{} percent, while adjusted EBITDA rose
\NumFreeUpworkEbitdaGrowthPct{} percent and the firm reported rising AI-related work
\citep{freelance_upwork_q2_2026}. The adjusted share-price comparison in
\cref{fig:app-freelance} shows that both firms' shares also fell, with much of the
decline beginning before coding agents arrived. These marketplaces include many forms
of work beyond software. Their results are a warning about some paid tasks, not proof
that customers have stopped buying software.

Stack Overflow shows how quickly such a picture can change. Monthly questions fell from
\NumSoBaselineTwentyTwoMonthly{} in 2022 to \NumSoLastMonthQuestions{} in
\NumSoLastMonth{} (\cref{fig:so}). The decline started earlier, with a \NumSoPreChatGPTDeclinePct{} percent fall between 2017 and 2022, so not all of it is due to AI.
The sharper later decline shows how quickly use can change, but this series does not
isolate AI's effect from the earlier trend.

\subsection{The warning: when benchmark tasks grow longer}
\label{sec:res-projection}
\Cref{fig:projection} extrapolates the measured capability trend.
To assess how an earlier trend compared with subsequent progress, we
fit the trend using only frontier models released before the
coding-agent era. At 50 percent success, all
\NumProjBacktestAboveN{} later frontier
models had task horizons above those predicted by the earlier trend,
with a median observed-to-predicted ratio of
\NumProjBacktestMedianRatio{} (\cref{fig:app-backtest}).
The earlier trend therefore understated subsequent benchmark progress.
This result does not establish that progress will continue to
accelerate. Model selection, changes in the task mix and benchmark
measurement may also affect the observed trend.

\datafigure[!htb]{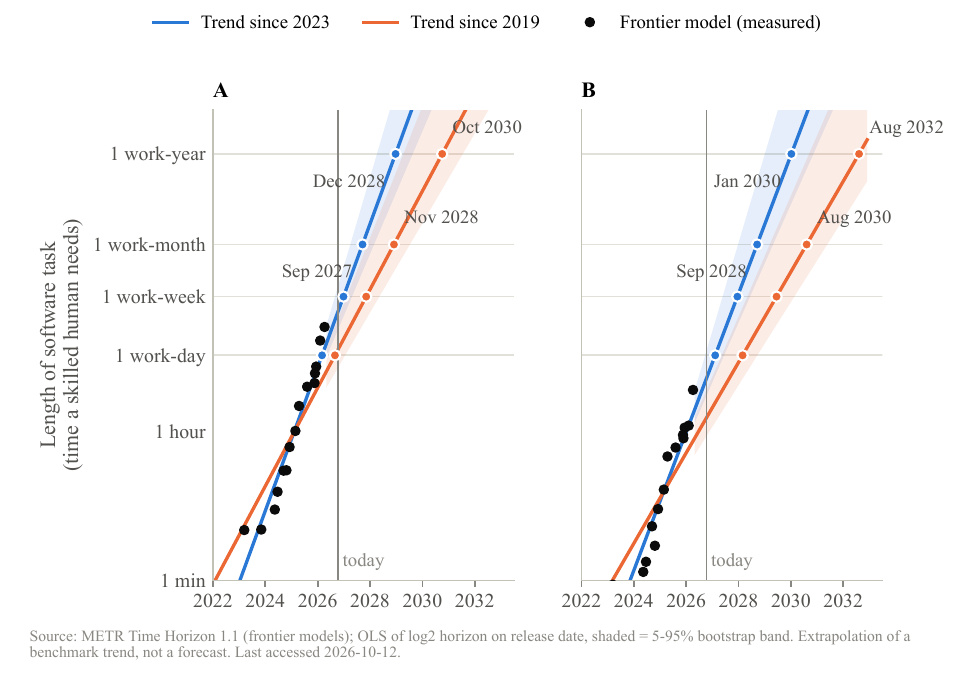}{\linewidth}
  {Task-horizon extrapolation: (A) \NumProjSuccessFifty{}\% success; (B) \NumProjSuccessEighty{}\% success; \NumProjBandLowPercentile{} to \NumProjBandHighPercentile{} percent bands.}
  {fig:projection}

If the more recent trend continues, agents would reach
\NumProjSuccessFifty{} percent success on benchmark tasks that would take
a skilled human a work-week around \NumProjFiftyFastWorkweekDate{}.
The corresponding work-month threshold would be reached around
\NumProjFiftyFastWorkmonthDate{} (interval:
\NumProjFiftyFastWorkmonthLow{} to \NumProjFiftyFastWorkmonthHigh{}).
Under the slower, longer-term trend, these thresholds would be reached
in \NumProjFiftySlowWorkweekDate{} and
\NumProjFiftySlowWorkmonthDate{}, respectively. Requiring
\NumProjSuccessEighty{} percent success moves the projected dates
later: the two trend fits place the work-month threshold between
\NumProjEightyFastWorkmonthDate{} and
\NumProjEightySlowWorkmonthDate{}, and the work-year threshold between
\NumProjEightyFastWorkyearDate{} and
\NumProjEightySlowWorkyearDate{}. These dates extrapolate benchmark
performance; they do not establish when agents could reliably build
and maintain complete software products.

These dates apply to performance on a specific set of benchmark tasks.
A work-month describes how long a skilled human would need to complete
a task, not how long an agent would run autonomously. Completing such
a task also does not establish that the resulting software can be
deployed and maintained reliably. The Craft suite illustrates how
quickly agents and contributors can create large codebases at an early
stage of development. Whether these become dependable replacements
for paid software depends on verification, integration and continued
maintenance. The projected dates are an extrapolation, not a forecast.
Future progress may differ as costs, available data and technical
constraints change. The measured series ends with models released in
\NumCapMetrLastRelease{}, so the projection does not incorporate
comparable task-horizon measurements for newer models.

\section{The Next Decade: When Building Changes Buying}
\label{sec:scenarios}

Consider a department within a company that pays for a screen recorder because it
needs to capture, trim and share clips. The department knows its workflow and can
assign a small team to build those functions with agents and existing components.
It does not need to reproduce the vendor's complete product or serve the vendor's
other customers. It needs a dependable tool for its own users. The same reasoning
applies to file archiving, document viewing and other bounded office workflows.
If the internal tool meets those needs at a lower total cost, the organization
could reduce its purchased seats or forgo a new contract. It could still pay for
model access, compute and upkeep. The cost comparison in \cref{sec:framework}
specifies the conditions for that choice.

Customers have other routes. A general AI subscription might handle a writing task
that once required a separate product, without anyone building an application. An
agent might instead create a private file utility or internal workflow that never
appears on GitHub. A customer could also adopt a public alternative and use an agent
to adapt it. Each route could reduce demand for a narrow application, but our public
repository and install data see only part of the process. Spending could move from
application subscriptions to model access, hosting and maintenance rather than
disappear.

Existing software may shorten the path. Published source can be reused when its
license permits. For software without source, the REA project documents tools that
let an agent inspect application behavior and binaries, then use the findings to
implement a similar feature \citep{rea_repository}. This shows that the tools
exist, not that they reproduce a whole product or remove legal and technical limits.
Agents can also inspect a visible interface. Current computer-use tools give models
screenshots and let them act through a mouse and keyboard
\citep{openai2026computeruse,anthropic2026computeruse}. A customer with access to an
application could have an agent try its features, observe the responses and attempt
to implement a similar workflow. Models can also examine product demonstrations in
video \citep{google2026videounderstanding}. These routes reveal observable behavior,
but they do not prove that an agent can recover the original code or deliver a
dependable replacement from a brief instruction.

Domain expertise could extend this route beyond simple office utilities. A
scientific or engineering team may know the physical model, input data, boundary
conditions and reference calculations for a research task. It could direct agents
to implement that workflow around existing numerical libraries, then check outputs
against known solutions. The commercial comparison would concern the functions
used in that research, rather than every feature of a general analysis package.
This is a proposed route to customer building. The scientific case discussed in
\cref{sec:background} illustrates both the role of specialist judgment and the
difficulty of obtaining trustworthy output.

Building need not be fully autonomous to change this choice. A domain expert can
set requirements, inspect intermediate results and decide when a tool is ready for
use, while agents carry out much of the implementation. HAS-Bench finds that the
benefits of participation depend on how it is organized
\citep{wu2026hasbench}. Its experiments simulate human participants, so they do not
establish the gains real professional teams would obtain. The economic test is
whether the combined human and agent effort costs less than buying an acceptable
tool. Human oversight belongs in that comparison rather than disqualifying the
build option.

A report published by Airfocus by Lucid describes a related organizational choice.
Lucid's product team evaluated commercial tools and built Archie, an internal tool
for account research using the team's own data
\citep{airfocus2026archie}. The account also describes usage costs, code review,
security work and uncertain maintenance. It illustrates a reported decision to
build instead of buy, not an independently audited cancellation or a rate of
replacement. These operational demands are part of the mechanism the trend data
leave unresolved.

The projection in \cref{fig:projection} supplies a conditional capability clock. At
\NumProjSuccessEighty{} percent success, the faster fit reaches tasks that would
take a skilled human a work-month in \NumProjEightyFastWorkmonthDate{}, and the slower
fit in \NumProjEightySlowWorkmonthDate{}. These are extrapolated dates for tasks in a fixed
suite, not dates when customers cancel subscriptions. A product combines many tasks
and must keep working after release. The projection cannot measure the time needed
for that transition. Recent models appear in other benchmarks
(\cref{fig:metr,fig:vibe-code,fig:app-swe}), but those scores cannot extend the METR horizon fit.
The first effect on vendors may instead be a customer's credible option to
build, which can put pressure on price or packaging before a sale is lost.

The market consequence depends on how far agents reduce the work after initial
implementation. An early route is a customer with domain expertise and technical
support directing agents through a bounded workflow. If agents also make testing,
deployment, integration and repair less demanding, smaller internal teams could
operate more complex tools. A large organization could then obtain more software
from a small group responsible for its specific needs. Multiple users, shared data
and continuing security work make that a harder test than producing a prototype.

\Needspace{5\baselineskip}
The broad extension of this mechanism reaches general computer users. If agents
can turn ordinary requests into suitable specifications, check their own work and
keep tools running, a person who can operate a computer could commission a private
tool without learning a programming language. This possibility gives the warning
its longer horizon. The present developer and benchmark data do not establish when
that level of assistance will be dependable or how widely it will be adopted.

\Cref{tab:scenarios} separates the ways spending could move. The routes can coexist;
they are conditional mechanisms, not estimated probabilities.

\begin{table}[H]
\centering
\small
\caption{How AI could change demand for applications.}
\label{tab:scenarios}
\begin{tabularx}{\linewidth}{@{}>{\raggedright\arraybackslash}p{0.22\linewidth} >{\raggedright\arraybackslash}X >{\raggedright\arraybackslash}X@{}}
\toprule
Route & Customer choice & What to measure \\
\midrule
AI subscription & Uses an assistant for a narrow task & Paid cancellations tied to assistant use \\
Private tool & An organization or professional buyer builds a tailored workflow & Reduced purchases, sustained use, oversight and total cost \\
Public alternative & Adopts or adapts a shared project & Sustained use and buyer reports of paid replacement \\
Vendor adaptation & Keeps buying for reliability, data, support or distribution & Renewals, prices and packaging among exposed vendors \\
\bottomrule
\end{tabularx}
\end{table}

By \NumProjCheckpointEarlyYear{}, comparable surveys should test whether the share
of in-house builders replacing paid vendors has risen above the
\NumSoSurveyReplacingVendors{} percent baseline. Buyer reports should distinguish a
private build from a general assistant replacing a task. In a narrow category,
sustained use of an alternative and evidence of reduced purchases would support
the customer-building route. Studies should distinguish organizations from
individual professionals and record the technical support each build requires.
More repositories alone would not establish that change.

By \NumProjCheckpointMiddleYear{}, exposed vendors should show pressure on seats,
prices, packaging or revenue if customers can credibly build or switch. These
changes should be compared with less exposed products and with broader market
conditions. The national investment series may miss an early private-build shift:
it values in-house software by developer pay and does not track canceled
application seats. Broader participation should also be tested directly: can users
with less programming experience specify, validate and operate useful tools with
less technical assistance? If alternatives multiply but renewals remain strong, the value of
maintenance, trust and integration is larger than the simple build story assumes.

Near-total disappearance of paid application software by
\NumProjCheckpointLongYear{} remains an extreme hypothesis, not our forecast. It
would require dependable replacement across many workflows, including those with
complex data, collaboration and legal responsibility. If agent capability rises
while direct measures of paid replacement stay flat through these checkpoints, the
warning is wrong in its strong form. The result would tell us how much of a
customer's payment was for things beyond writing code.

\section{Where Software Value May Move}
\label{sec:implications}

The first change visible in our data is a wider set of ways to obtain software.
Repositories describing themselves as alternatives have grown relative to all new
repositories (\cref{fig:altrate}), and the product search found
\NumAltIncluded{} projects that passed its broad screening rule
(\cref{fig:broad}). That count is an upper bound until validation. Yet the
national share of software built in-house has not risen (\cref{fig:bea}), and only
\NumSoSurveyReplacingVendors{} percent of surveyed in-house builders said they were
replacing a paid vendor. These measures locate the present tension: candidate
substitutes are more numerous, while a broad change in buying remains unobserved. A
vendor should therefore examine the workflows customers can specify and check for
themselves. Narrow tools may face a credible build option sooner. Shared data,
integrations, support and legal responsibility may retain value where customers
cannot supply them at a lower total cost. The evidence does not yet tell us how much
those features protect prices or renewals.

For organizational buyers, domain knowledge and access to data may become more
useful inputs to internal development as programming becomes easier to obtain.
A scientist who can specify a calculation or an operations team that understands
its business rules can help direct and evaluate a build. The potential change is
in the technical effort needed to turn that knowledge into a maintained tool.
Vendors may therefore face competition from small teams inside their own customer
organizations, even when those teams have no intention of selling software.

The Craft suite shows why implementation speed alone cannot settle that comparison.
It produced \NumCraftApps{} applications in \NumCraftHistoryDays{} days of public
history, but its own roadmaps rate them a median of \NumCraftReadyMedian{} percent
ready for real work (\cref{fig:craft,tab:craft}). A buyer considering a private tool
must pay for verification, migration, hosting, security, failures and later changes,
as well as the first build (\cref{sec:framework}). Developers' work in specifying
acceptance standards, testing, connecting and maintaining systems therefore matters
to the choice between building and buying. Agents may eventually perform more of
that work too; our evidence does not measure when they will do it reliably.

Some vendors may try to slow imitation by withholding technical guides, publishing
only video tutorials, or showing an instruction inside the product that tells agents
not to inspect it. We have no evidence that these steps prevent substitution.
Computer-use agents can observe visible behavior, and models can process video
\citep{openai2026computeruse,google2026videounderstanding}. From the agent's perspective, an instruction displayed by the product is untrusted content. It may distract
some agents, but provider guidance tells them not to treat such text as authority
over the customer's request \citep{openai2026computeruse,anthropic2026browseruse}.
Withholding documentation can also make the product harder for customers to learn
and integrate. These tactics might add friction; their effect on retention has not
been measured.

The current market need not be a simple transfer from a paid product to a free one.
In the Homebrew pilot, Bruno's installs reached \NumTenBrewBrunoPctPostman{} percent
of Postman's, while Postman's own installs also rose (\cref{fig:homebrew}). Those
counts cover only part of the market and do not measure canceled subscriptions.
The thin contributor base of newer projects in the narrow pilot adds a second
uncertainty: an alternative can be easy to start and hard to sustain. If customers
do replace application subscriptions, they may still pay for model access, hosting,
maintenance or a vendor that supplies trusted service. The next round of evidence should ask which of those costs customers take on, and which firms earn
the resulting revenue.

\section{Limits of the Evidence}
\label{sec:validity}

The data do not link an identified customer's internal build to a verified
reduction in purchases. Repositories and packages record creation, while
stars, downloads and installs indicate attention or acquisition. The
organizational survey reports decisions to forgo purchases, but does not
verify the resulting tools, their continued use or the spending displaced
\citep{customer_building_mckinsey2026}. Public sources miss most private
tools, and national accounts value in-house software by developer pay
rather than output. These sources therefore cannot establish the rate
of customer substitution.

The observed populations and measures also limit generalization.
GitHub contributors, developer survey respondents and Homebrew users
do not represent all organizational or professional buyers. The company
and scientific cases are selected illustrations. Deleted repositories,
package spam, repeated installs and inconsistent AI attribution affect
the public counts. The product sample is small, and some workflow
matches rely on keywords, making the broad alternative counts an upper
bound pending independent validation. AI attribution records credited
commits rather than the share of code written by AI.

The timing comparisons cannot isolate the effect of coding agents.
Changes in hiring, interest rates and US tax treatment of software
development provide competing explanations. Comparing break dates and
normalizing activity cannot remove all such influences. Freelance
platform results and share prices provide additional context, but
cannot establish AI's effect on software demand.

The capability projections introduce further uncertainty. METR's series
ends with models released in \NumCapMetrLastRelease{} and measures
benchmark tasks rather than maintained products. The backtest uses
selected frontier models, and the bootstrap omits uncertainty from task
sampling. Newer benchmarks use different scales and cannot extend the
METR fit
\citep{agent_capability_cost_epoch_eci,agent_capability_cost_epoch_frontierswe,customer_building_vals2026}.
Their scores also do not establish security or maintenance quality.
The projected dates are an extrapolation, not a forecast of customer
switching. A direct test would follow customers through building,
verification, maintenance and subsequent purchasing decisions.
\section{Conclusion}
\label{sec:conclusion}

AI coding agents could change the economic relationship between software
companies and their customers. Organizations may already have the domain
knowledge, data and acceptance criteria needed to specify useful tools.
Agents and reusable code could help small teams turn those resources into
software tailored to their needs. A credible option to build could
therefore affect purchasing decisions before a subscription is canceled,
changing what customers are willing to pay and how vendors compete.

The evidence shows growing capacity to build, alongside mixed evidence
of changes in purchasing. Agent benchmark capability has improved, the
cost of a fixed coding result has fallen, and public software creation
rose sharply around the arrival of coding agents. Alternatives to paid
products are appearing, although some ambitious projects remain
unfinished. Available market indicators still show strong purchasing:
packaged software gained investment share, tracked commercial installs
remain strong, and most surveyed in-house builders report purposes other
than replacing vendors. An organizational survey nevertheless records
decisions to forgo purchases in favor of building. These reports suggest
that the mechanism is affecting some buying decisions, but leave the
spending displaced and the reliability of the resulting tools unresolved.
The warning is that customers' capacity to build may expand before its
market effects become widely visible.

Whether this capacity leads to sustained substitution depends on the
total cost of producing, verifying, integrating and maintaining software
that meets customers' needs. The software substitution frontier frames
this comparison at the customer workflow level. The scenarios
identify observable tests of whether organizational teams and domain
experts can replace purchases, and whether that ability extends to
broader users as agents take on more of the work. The capability dates
are an extrapolation, not a forecast of adoption. Following customers
through building, sustained use and subsequent purchasing decisions
would establish whether the proposed shift is occurring. For software
companies, the central question is how much value they can continue to
provide when their customers have a credible option to build.

\section*{Data and code availability}
The code, notebooks, and data supporting this study are publicly available on
GitHub at \url{https://github.com/riponcm/The-Customer-Is-the-Builder}.
An archived release is available on Zenodo at \url{https://doi.org/10.5281/zenodo.23524847}.

\Needspace{7\baselineskip}
\section*{Disclosures}
The first author developed parts of the data collection and analysis code with assistance from AI coding agents.
We used PROJECTMEM as a project memory tool during notebook development and data analysis
\citep{malo2026projectmem}. The first and second authors checked all results.
The first author has no financial relationship with any company or project discussed.
\bibliographystyle{plainnat}
\begingroup
\raggedright
\setlength{\bibsep}{4pt}
\interlinepenalty=10000
\bibliography{references,data_sources}
\endgroup

\clearpage
\appendix
\section{Additional Figures and Tables}
\label{app:figures}

\begin{table}[H]
\centering
\scriptsize
\caption{Interrupted time series of new packages and downloads.}
\label{tab:its-packages}
\resizebox{\linewidth}{!}{
\begin{tabular}{ll r@{\,}l r@{\,}l r@{\,}l r}
\toprule
\multicolumn{9}{l}{\textit{Panel A: break at Dec 2022}} \\
Ecosystem & Series & \multicolumn{2}{c}{Pre-trend (\%/yr)} & \multicolumn{2}{c}{Level, Dec 22 (\%)} & \multicolumn{2}{c}{Slope chg., Dec 22 (\%/yr)} & $N$ \\
\midrule
PyPI & new packages & $+25.8^{***}$ & $(1.8)$ & $-35.5^{**}$ & $(15.2)$ & $-14.0^{***}$ & $(3.8)$ & 129 \\
PyPI & filtered & $+21.7^{***}$ & $(1.5)$ & $-26.1^{**}$ & $(11.5)$ & $-5.5^{**}$ & $(2.6)$ & 126 \\
crates.io & new packages & $+33.5^{***}$ & $(1.7)$ & $-50.5^{**}$ & $(23.9)$ & $+15.6$ & $(11.7)$ & 129 \\
crates.io & filtered & $+31.9^{***}$ & $(1.6)$ & $-40.4^{**}$ & $(19.8)$ & $+9.0$ & $(10.6)$ & 126 \\
RubyGems & new packages & $-16.5^{***}$ & $(0.7)$ & $-11.7$ & $(17.1)$ & $+41.6^{***}$ & $(11.3)$ & 129 \\
RubyGems & filtered & $-17.3^{***}$ & $(0.6)$ & $-19.0$ & $(14.5)$ & $+37.9^{***}$ & $(8.2)$ & 129 \\
Packagist & net additions & $-0.2$ & $(1.8)$ & $-34.7^{***}$ & $(13.4)$ & $+11.9^{**}$ & $(5.4)$ & 129 \\
\addlinespace
npm & downloads & $+57.5^{***}$ & $(3.9)$ & $-59.8^{***}$ & $(19.4)$ & $-22.0^{***}$ & $(6.3)$ & 116 \\
PyPI & downloads & $+58.3^{***}$ & $(1.4)$ & $-31.8^{***}$ & $(6.9)$ & $+3.1$ & $(2.5)$ & 97 \\
Packagist & installs & $+39.0^{***}$ & $(0.9)$ & $-33.1^{***}$ & $(4.7)$ & $-16.8^{***}$ & $(2.5)$ & 117 \\
\midrule
\multicolumn{9}{l}{\textit{Panel B: breaks at Dec 2022 and Feb 2025}} \\
Ecosystem & Series & \multicolumn{2}{c}{Slope chg., Dec 22 (\%/yr)} & \multicolumn{2}{c}{Level, Feb 25 (\%)} & \multicolumn{2}{c}{Slope chg., Feb 25 (\%/yr)} & $N$ \\
\midrule
PyPI & new packages & $-28.2^{***}$ & $(6.0)$ & $+17.3^{*}$ & $(10.4)$ & $+19.5^{***}$ & $(7.6)$ & 129 \\
PyPI & filtered & $-15.5^{***}$ & $(3.0)$ & $+10.4^{***}$ & $(4.0)$ & $+17.9^{***}$ & $(3.8)$ & 126 \\
crates.io & new packages & $-21.9^{***}$ & $(2.4)$ & $-24.1^{***}$ & $(8.9)$ & $+121.6^{***}$ & $(9.5)$ & 129 \\
crates.io & filtered & $-19.3^{***}$ & $(2.4)$ & $-27.1^{***}$ & $(10.3)$ & $+119.6^{***}$ & $(14.3)$ & 126 \\
RubyGems & new packages & $+27.4^{*}$ & $(15.3)$ & $-81.3^{***}$ & $(31.0)$ & $+118.2^{***}$ & $(15.4)$ & 129 \\
RubyGems & filtered & $+10.4^{***}$ & $(1.8)$ & $-13.4^{***}$ & $(4.5)$ & $+84.9^{***}$ & $(4.9)$ & 129 \\
Packagist & net additions & $-5.8^{***}$ & $(2.2)$ & $-9.2^{*}$ & $(5.1)$ & $+55.1^{***}$ & $(6.4)$ & 129 \\
\addlinespace
npm & downloads & $-41.2^{***}$ & $(4.3)$ & $-8.0$ & $(5.3)$ & $+59.8^{***}$ & $(7.1)$ & 116 \\
PyPI & downloads & $-1.8$ & $(2.1)$ & $-11.0^{***}$ & $(3.2)$ & $+25.1^{***}$ & $(3.5)$ & 97 \\
Packagist & installs & $-16.5^{***}$ & $(2.1)$ & $-13.5^{*}$ & $(7.6)$ & $+12.6$ & $(7.8)$ & 117 \\
\bottomrule
\end{tabular}
}
\par\smallskip
\parbox{\linewidth}{\scriptsize Panel A has one break at December 2022. Panel B adds a
break at February 2025. Coefficients in log points times 100. Standard errors allow for
autocorrelation (12 lags). Month-of-year effects included.}
\end{table}

\begin{table}[H]
\centering
\small
\caption{Package trend changes across nearby break dates.}
\label{tab:break-sensitivity}
\resizebox{\linewidth}{!}{
\begin{tabular}{@{}lrrl@{}}
\toprule
Series & February slope change & Range across breaks & Lowest AIC break \\
\midrule
PyPI screened & $+17.9$ & $+12.6$ to $+25.6$ & Aug 2024 \\
crates.io screened & $+119.6$ & $+82.0$ to $+177.6$ & Aug 2025 \\
RubyGems screened & $+84.9$ & $+66.1$ to $+92.3$ & May 2025 \\
Packagist net additions & $+55.1$ & $+44.6$ to $+73.1$ & Aug 2025 \\
\bottomrule
\end{tabular}
}
\par\smallskip
\parbox{\linewidth}{\footnotesize Annualized change in log points after the
second break. The range moves that break monthly from
\NumPkgSensitivityFirstBreak{} to \NumPkgSensitivityLastBreak{};
the ChatGPT break remains fixed. Lowest AIC identifies the best fit among these
dates, not a causal event date. Source: package registries in \cref{tab:sources}.}
\end{table}

\begin{table}[H]
\centering
\scriptsize
\caption{The Craft suite at collection.}
\label{tab:craft}
\resizebox{\linewidth}{!}{
\begin{tabular}{@{}l l l r r r r r r r l@{}}
\toprule
App & Incumbent & Created & Commits & LOC (k) & Contrib. & AI (\%) & CI (\%) & Stars & Dl.\ (k) & Ready (\%)$^{a}$ \\
\midrule
PhotoCraft & Adobe Photoshop & Sep 30 & 1,616 & 443 & 225 & 67 & 59 & 49,277 & 886 & 45 \\
PdfCraft & Adobe Acrobat Pro & Sep 30 & 825 & 195 & 149 & 72 & 48 & 10,029 & 532 & 49 \\
FilmCraft & Adobe Premiere Pro & Sep 30 & 1,756 & 372 & 88 & 48 & 68 & 10,496 & 493 & 56 \\
VectorCraft & Adobe Illustrator & Sep 30 & 2,201 & 371 & 88 & 50 & 87 & 7,290 & 442 & 62 \\
LightCraft & Adobe Lightroom & Sep 30 & 1,938 & 209 & 97 & 69 & 59 & 10,441 & 416 & 61 \\
DesignCraft & Adobe InDesign & Oct 1 & 494 & 123 & 33 & 68 & 69 & 3,114 & 291 & 47 \\
EffectCraft & Adobe After Effects & Oct 1 & 1,306 & 320 & 51 & 67 & 70 & 5,301 & 367 & 45 \\
GridCraft & Microsoft Excel & Oct 7 & 599 & 114 & 40 & 33 & 47 & 3,175 & 151 & 50 \\
CADCraft & Autodesk AutoCAD & Oct 7 & 356 & 146 & 29 & 83 & 57 & 2,870 & 130 & 38 \\
DeckCraft & Microsoft PowerPoint & Oct 7 & 62 & 111 & 15 & 87 & 76 & 2,145 & 129 & 55 \\
SoundCraft & Avid Pro Tools Ultimate & Oct 7 & 305 & 73 & 20 & 44 & 54 & 1,932 & 113 & 40 \\
WordCraft & Microsoft Word & Oct 7 & 302 & 155 & 45 & 88 & 70 & 6,862 & 195 & 60 \\
\midrule
\textbf{Suite} & 12 apps &  & 11,760 & 2,634 &  & 61 &  & 112,932 & 4,145 &  \\
\bottomrule
\end{tabular}
}
\par\smallskip
\parbox{\linewidth}{\scriptsize Source: project repositories
\citep{craft_github_org}. AI is the share of commits credited to an AI tool.
CI is the share of successful GitHub Actions runs. Downloads include
automatic updates and other assets. $^{a}$ Readiness is the project's own estimate.}
\end{table}

\datafigure[H]{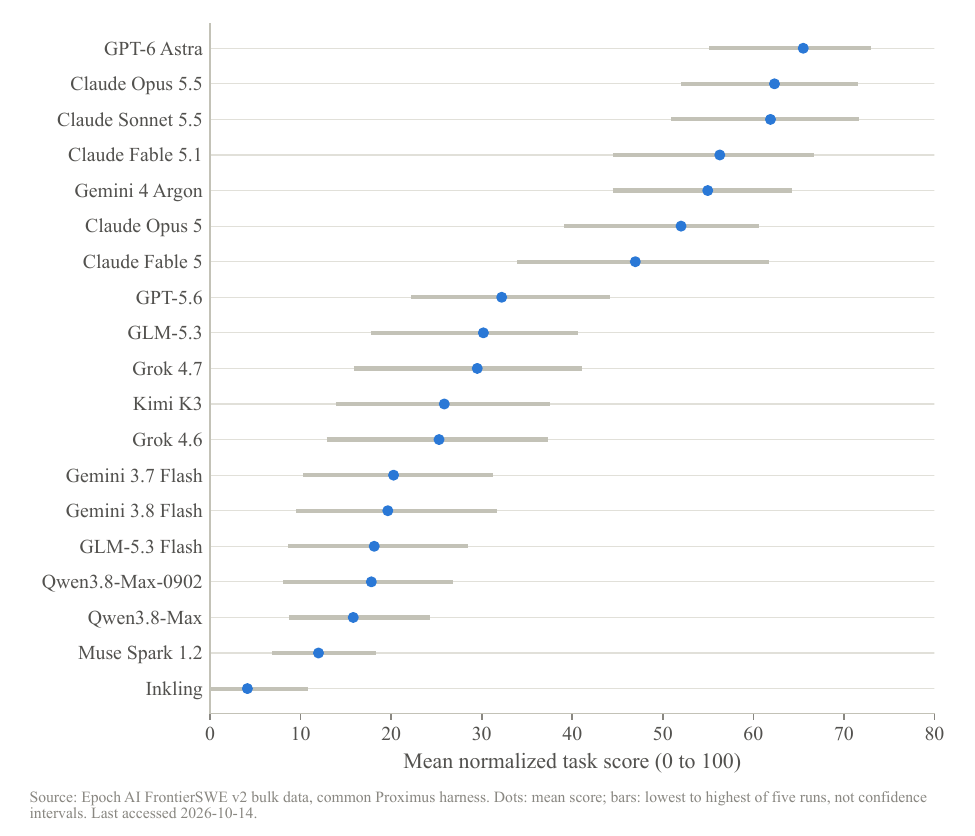}{\linewidth}
  {Recent models on the FrontierSWE coding benchmark.}
  {fig:app-swe}

\datafigure{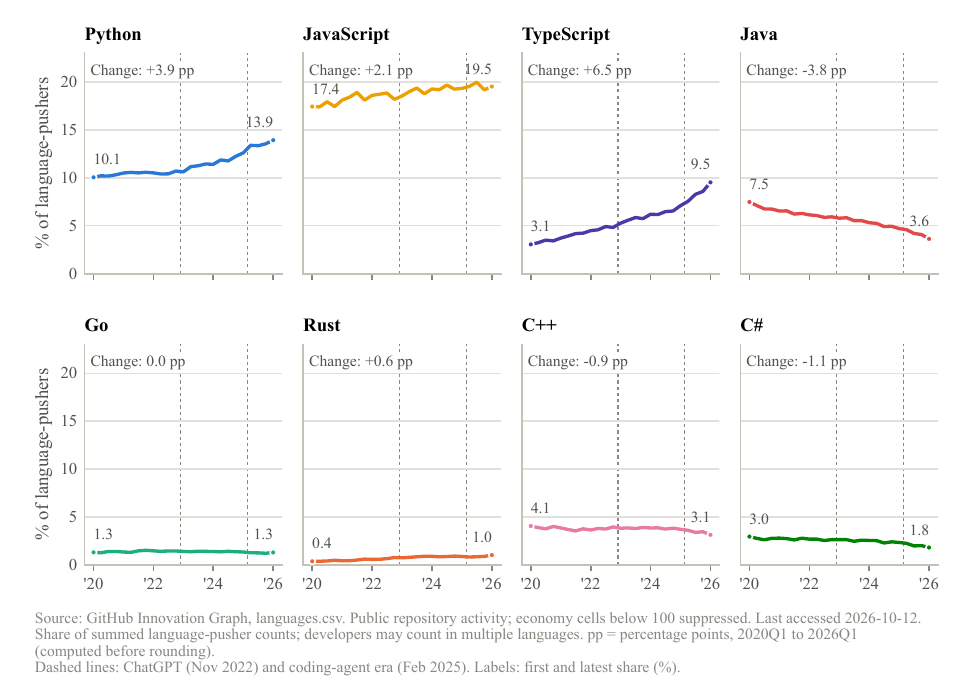}{\linewidth}
  {Share of developers pushing code, by language.}
  {fig:app-lang}

\datafigure{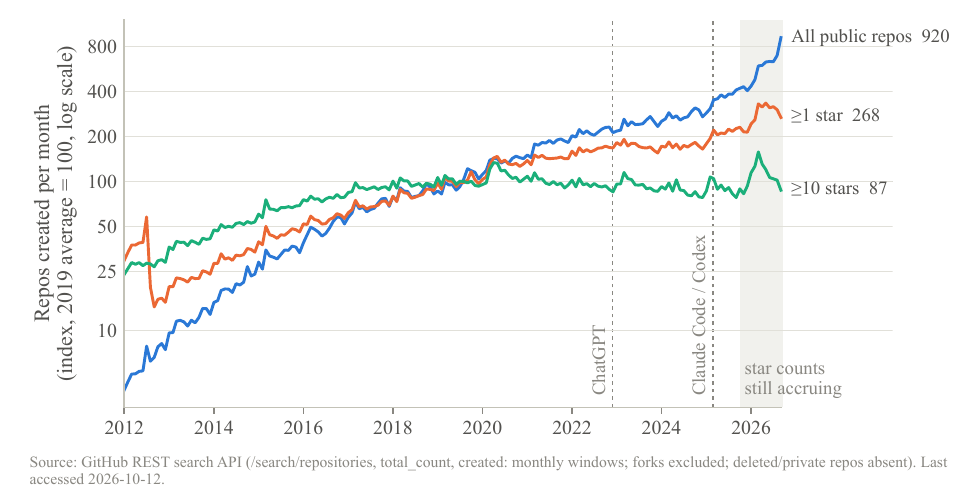}{\linewidth}
  {New public GitHub repositories per month.}
  {fig:app-repos}

\datafigure{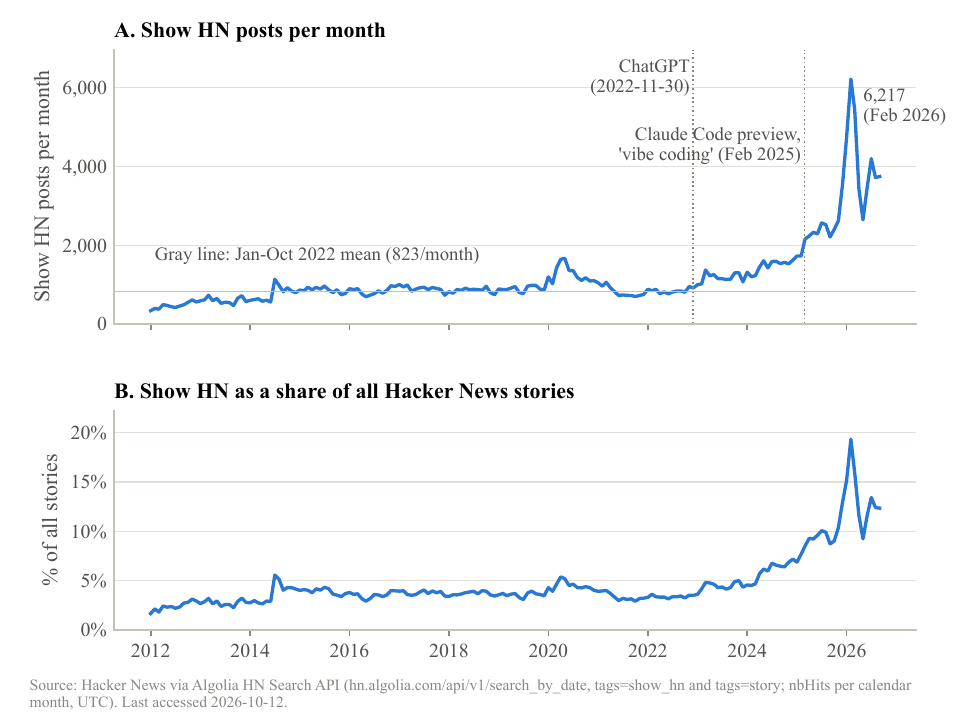}{\linewidth}
  {Show HN posts per month.}
  {fig:app-showhn}

\datafigure{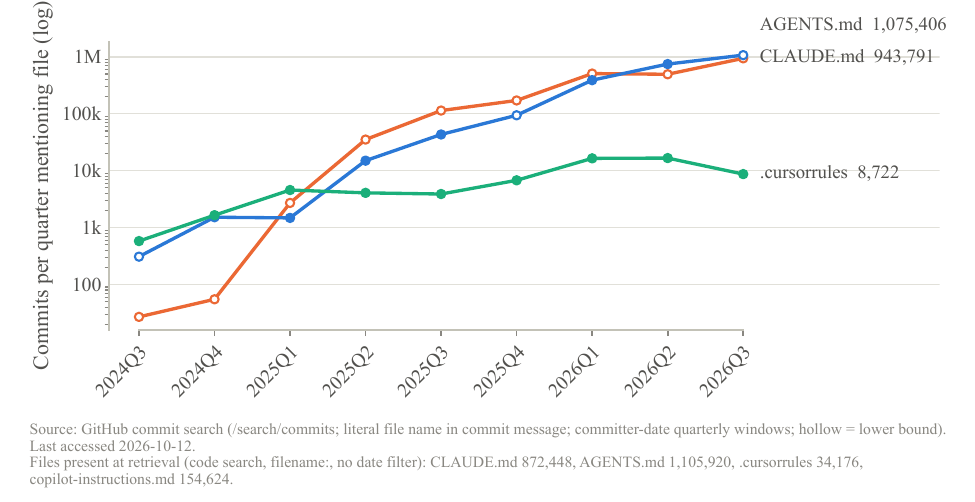}{\linewidth}
  {Commits that mention an agent instruction file.}
  {fig:app-instr}

\datafigure{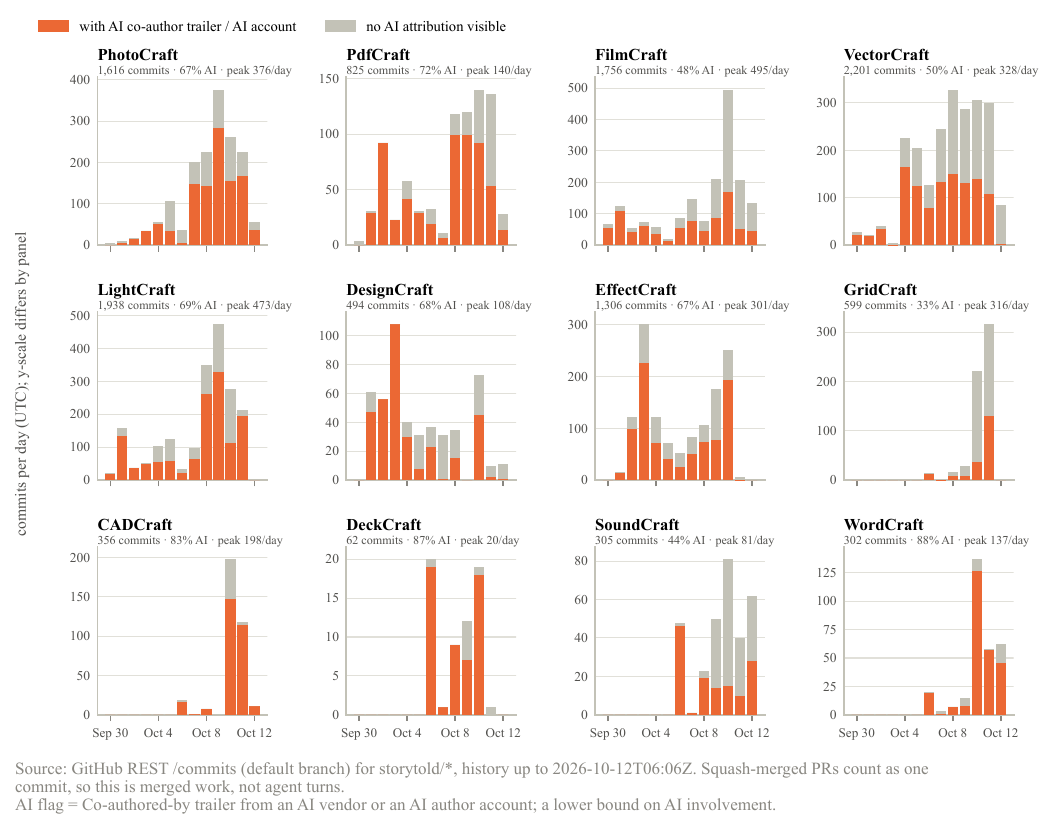}{\linewidth}
  {Craft suite commits per day, with AI attribution.}
  {fig:app-craftcommits}

\datafigure{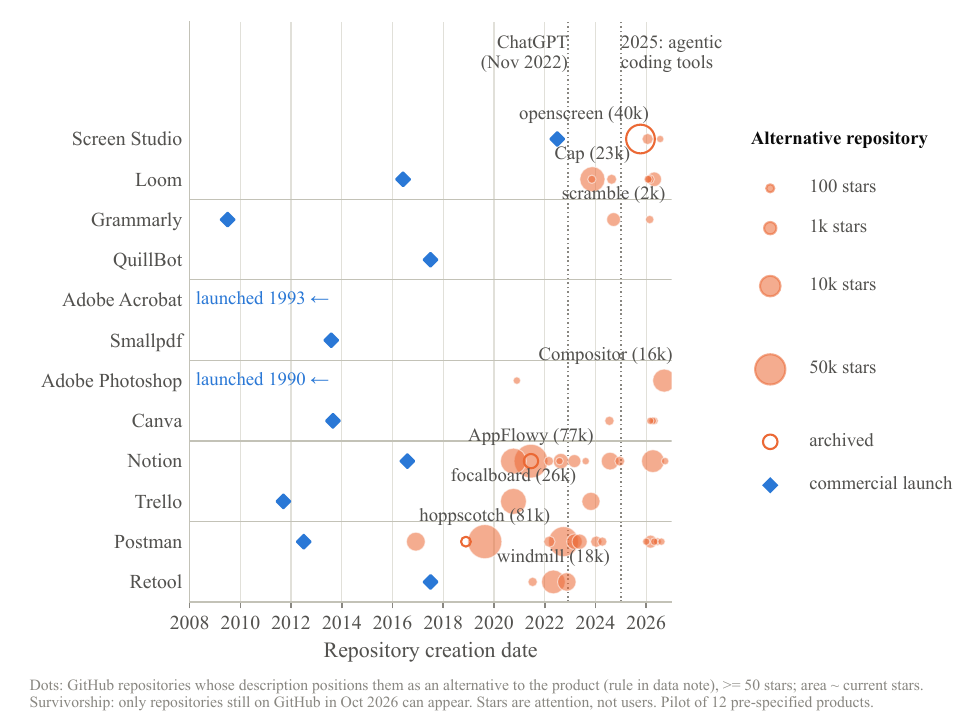}{\linewidth}
  {Alternatives that name the product, by creation date (narrow rule).}
  {fig:app-narrow}

\datafigure{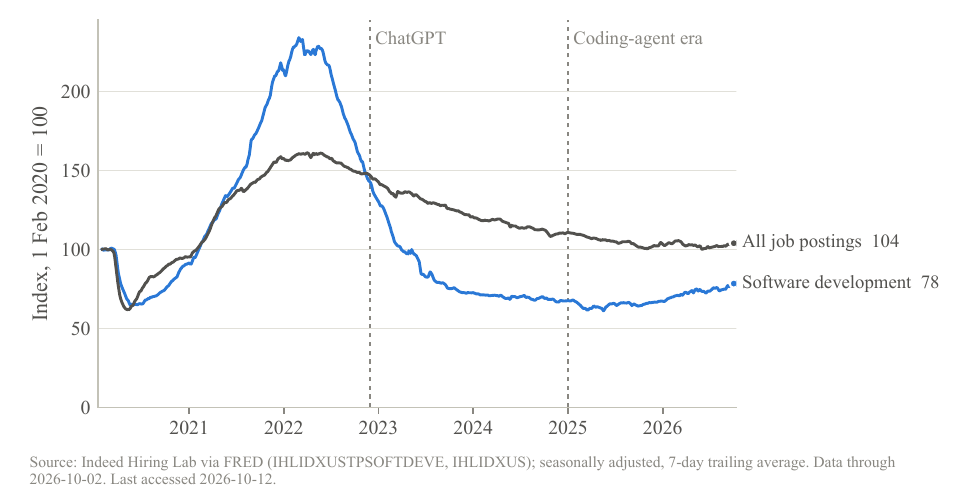}{\linewidth}
  {Job postings for software development and for all jobs (Indeed).}
  {fig:app-jobs}

\datafigure{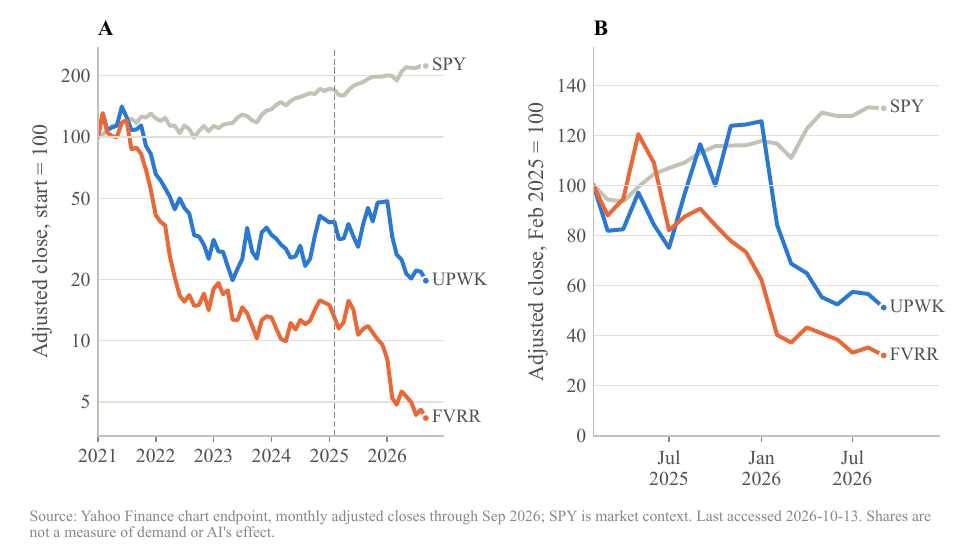}{\linewidth}
  {Freelance platform shares versus SPY: (A) full period; (B) coding-agent era.}
  {fig:app-freelance}

\datafigure{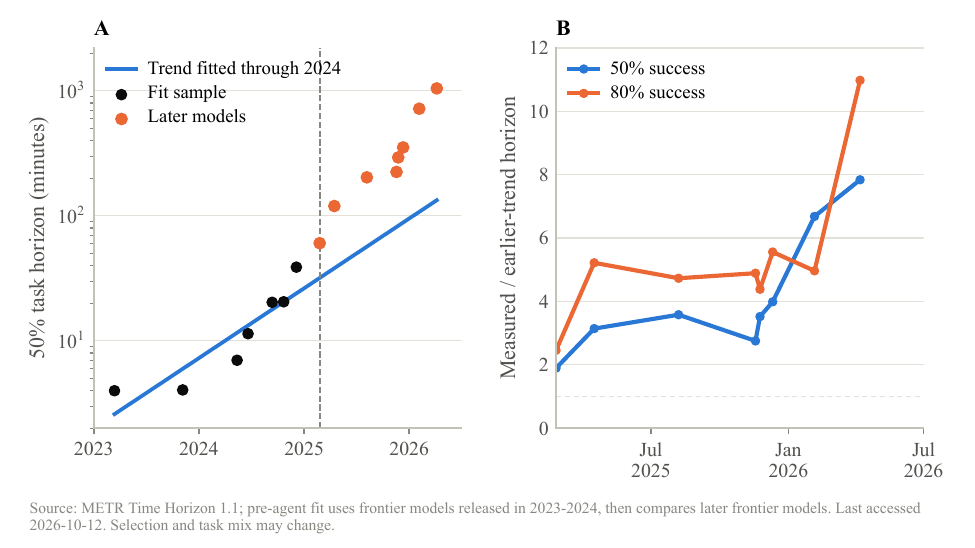}{\linewidth}
  {Capability backtest: (A) earlier trend versus later models; (B) observed-to-predicted horizon ratios.}
  {fig:app-backtest}

\end{document}